\documentclass[showpacs,prd,10pt,twocolumn,superscriptaddress,
nofootinbib,aps]{revtex4-1}

\usepackage[utf8]{inputenc}
\usepackage{ae}
\usepackage{xcolor}
\usepackage{amsmath}
\usepackage{amssymb}
\usepackage{upgreek}
\usepackage{amsfonts}
\usepackage{graphicx}
\usepackage{slashed}
\usepackage{siunitx}
\usepackage{hyperref}
\usepackage{cleveref}
\usepackage{easyReview}
\usepackage{listings}
\usepackage{placeins}
\usepackage{csquotes}
\newcommand{\ie}{{i.e.}, }
\newcommand{\eg}{{e.g.}, }

\newcommand{\vs}{{vs.} }
\newcommand{\cf}{{cf.} }

\renewcommand{\i}{\mathrm{i}}
\newcommand{\e}{\mathrm{e}}

\newcommand{\vect}[1]{{\boldsymbol{#1}}}
\newcommand{\diff}[2]{\mathop{}\!\mathrm{d}^{#1}#2\;}

\newcommand{\FFT}{ \operatorname{FFT}}
\newcommand{\iFFT}{ \operatorname{iFFT}}

\crefname{equation}{\text{Eq.}}{\text{Eqs.}}
\Crefname{equation}{\text{Equation}}{\text{Equations}}
\crefname{section}{\text{Sec.}}{\text{Secs.}}
\Crefname{section}{\text{Section}}{\text{Sections}}
\crefname{appendix}{\text{Appendix}}{\text{Appendixes}}
\Crefname{appendix}{\text{Appendix}}{\text{Appendixes}}
\crefname{table}{\text{Tab.}}{\text{Tabs.}}
\Crefname{table}{\text{Table}}{\text{Tables}}
\crefname{figure}{\text{Fig.}}{\text{Figs.}}
\Crefname{figure}{\text{Figure}}{\text{Figures}}
\crefname{lstlisting}{\text{List.}}{\text{Lists.}}
\Crefname{lstlisting}{\text{Listing}}{\text{Listings}}
\crefrangeformat{equation}{Eqs.~(#3#1#4)--(#5#2#6)}
\crefrangeformat{appendix}{Appendixes~(#3#1#4)--(#5#2#6)}

\makeatletter
\def\fps@figure{ht}
\makeatother

\allowdisplaybreaks

\definecolor{shiraz}{rgb}{0.6,0,0.4}

\usepackage{soul}

\hypersetup{
    colorlinks=True
}

\graphicspath{{./figures/}}

\begin{document}

\setlength{\unitlength}{1mm}

\title{Phase transition analogs in laser collisions with a dark-field setup}

\author{Holger Gies}
\email{holger.gies@uni-jena.de}
\affiliation{Theoretisch-Physikalisches Institut, Abbe Center of Photonics,
    Friedrich-Schiller-Universität Jena, Max-Wien-Platz 1, 07743 Jena, Germany}
\affiliation{Helmholtz-Institut Jena, Fröbelstieg 3, 07743 Jena, Germany}
\affiliation{GSI Helmholtzzentrum für Schwerionenforschung, Planckstr. 1,
    64291 Darmstadt, Germany}
\author{Felix Karbstein}
\email{felix.karbstein@uni-jena.de}
\affiliation{Theoretisch-Physikalisches Institut, Abbe Center of Photonics,
    Friedrich-Schiller-Universität Jena, Max-Wien-Platz 1, 07743 Jena, Germany}
\affiliation{Helmholtz-Institut Jena, Fröbelstieg 3, 07743 Jena, Germany}
\affiliation{GSI Helmholtzzentrum für Schwerionenforschung, Planckstr. 1,
    64291 Darmstadt, Germany}
\author{Lars Maiwald}
\email{lars.maiwald@uni-jena.de}
\affiliation{Theoretisch-Physikalisches Institut, Abbe Center of Photonics,
    Friedrich-Schiller-Universität Jena, Max-Wien-Platz 1, 07743 Jena, Germany}
\affiliation{Helmholtz-Institut Jena, Fröbelstieg 3, 07743 Jena, Germany}
\affiliation{GSI Helmholtzzentrum für Schwerionenforschung, Planckstr. 1,
    64291 Darmstadt, Germany}
\date{\today}

\begin{abstract}
    Laser pulse collisions are a promising tool for the investigation of
    light-by-light scattering phenomena induced by quantum vacuum fluctuations.
    Using the numerical code based on the vacuum emission picture and put
    forward in \cite{Blinne:2018nbd}, we observe a strong dependence of the
    signal features on the transverse profiles of the colliding laser
    pulses in the interaction region. For a probe beam tailored such as to
    feature an annular far-field profile and a pronounced on-axis focus peak
    counterpropagating a pump beam at zero impact parameter, the signal's main
    emission direction can undergo the analog of a phase transition with the
    beam-waist ratio of the pulses serving as a control parameter. Depending on
    the pump's beam profile, this phase transition can be first order (\eg for a
    pump with a flat-top far-field profile) or second order (\eg for a Gaussian
    pump). From the simulation data, we determine the critical point and extract
    the corresponding critical exponent for the second-order transition of the
    main emission direction of the signal in the far field. For this, we improve
    the performance of the above numerical code, using the phase transition
    analogs as an example to illustrate the capabilities and limitations of
    the code and current workflows.
\end{abstract}

\maketitle

\section{Introduction} \label{sec:intro}

Nonlinear interactions of electromagnetic fields in vacuum are a long-standing
prediction of quantum electrodynamics (QED)
\cite{Heisenberg:1936nmg,weisskopf1936}. For a precise and quantitative
verification of these interactions, the collision of macroscopically
controllable, ultra-intense laser pulses is a promising
experimental pathway in many laboratories worldwide. This could provide access
to the plethora of phenomena induced by quantum vacuum fluctuations of charged
matter degrees of freedom; see
\cite{dittrich2000,Dunne:2004nc,DiPiazza:2007prw,Battesti:2012hf,King:2015tba,Karbstein:2019oej,Fedotov:2022ely}
for reviews.

The direct route to experimental investigation is hampered by the diminutiveness
of the effect, as can be seen on the quantum level from the small photon-photon
cross section at low frequency
\cite{Euler:1935zz,Karplus:1950zza,Karplus:1950zz,DeTollis:1965vna}, requiring a
careful and delicate separation of the small signal from the expectably large
background. Many ideas and concepts have been put forward in recent years to
solve this problem; \cf \eg
\cite{Lundstrom:2005za,King:2010nka,King:2012aw,Karbstein:2015xra,Ataman:2018ucl,Aboushelbaya:2019ncg,Karbstein:2020gzg,Dumlu:2022olg,Karbstein:2022uwf,Sundqvist:2023hvw,Formanek:2023mkx,Macleod:2024jxl}.
A particularly promising idea is the use of tailored probe beams in order to
realize a dark-field setup, which combines the virtues of a geometrical
signal-to-background separation with the still sizable scattering rates in the
forward direction of the probe in a comparatively simple configuration involving
only two counterpropagating beams
\cite{Karbstein:2020gzg,Karbstein:2022uwf,Schutze:2024kzu}. A similar scheme has
already experimentally proved of value in high-harmonic generation
\cite{peatross1994,zepf1998} and has become a cornerstone of an intended
discovery experiment of vacuum birefringence at the HED-HIBEF beamline at the
European XFEL \cite{Ahmadiniaz:2024xob}.

While theoretical tools for the prediction of observables for realistic
laser pulses have also evolved substantially in the past years (see
\cite{Gonoskov:2021hwf,Fedotov:2022ely} and references therein), analytical
methods are often limited to validity regimes of necessary approximations. A
particularly efficient analytical modeling of realistic laser pulses is often
achieved by using paraxial beams together with the approximation of taking the
formal limit of an infinite Rayleigh range \cite{Gies:2017ygp,King:2018wtn}. This
approximation can, for instance, be quantitatively well controlled for
nonextremal focusing and sufficiently small pulse durations \cite{Karbstein:2021ldz}.

For general beams and collision geometries, powerful numerical methods are
highly desired. In addition to methods that aim at numerically solving the
QED-induced nonlinear Heisenberg-Euler-Maxwell equations in a general fashion
\cite{Grismayer:2016cqp,Lindner:2022alm,Lindner:2023pfl}, methods that are
designed to directly compute the relevant observables have turned out to be
particularly fruitful for the description of pulse collisions. In the present
work, we use the VacEm code \cite{Blinne:2018nbd} which is based on the vacuum
emission picture \cite{GALTSOV1971238,Karbstein:2014fva}. The latter effectively
reduces the computation of the quantum-induced signatures to performing a
space-time Fourier transformation of the field distribution in the interaction
region; the corresponding numerical task can be faced in various ways
\cite{Karbstein:2015xra,Sainte-Marie:2022efx,Berezin:2024fxt}. Specifically, the
VacEm code has been used in various studies requiring quantitative accuracy
\cite{Blinne:2018nbd,Karbstein:2019dxo,Schutze:2024kzu,Valialshchikov:2024svm}.

In the present work, we highlight and analyze a phenomenon in the quantum-vacuum
signatures of colliding pulses that is reminiscent of a phase transition in
statistical physics. We demonstrate that both its qualitative and quantitative
properties depend sensitively on the beam properties such that the use of an
accurate computational method is mandatory. More specifically, we use a
\textit{probe} beam with an annular far-field profile but a pronounced on-axis
peak in the focus colliding head on with a strong \textit{pump} pulse, and we study
the angular distribution of quantum-induced signal photons arising from the
interaction region. We confirm that the far-field emission characteristics of
the signal can be peaked either in the direction of the probe's beam axis or at
a finite angular offset \cite{Karbstein:2020gzg} depending on various details,
specifically the relative beam waists of the colliding pulses and the transverse
beam profiles.

Interestingly, our results can be phrased in the language of critical phenomena
with the main emission direction of the signal serving as an order parameter
that can undergo an apparent symmetry transition from on axis to off axis
depending on the pump properties serving as control parameters. As the
computation of observables in the critical region---specifically the location of
the phase transition, its order, and a corresponding critical exponent in the
case of a second-order transition---requires a substantial numerical accuracy, we
use the present investigation also as a motive to improve the VacEm code as
well as study its convergence with critical discretization parameters, as
initiated in \cite{maiwald2023}. We observe that the accurate modeling of
flat-top beam profiles, which are rather generic in experiments, can pose a
numerical challenge. Addressing this challenge, our code improvements
target the computational cost. More specifically, the improvements
significantly reduce the computation time, memory, and storage demand. The
extensive simulations of flat-top beams are thus made feasible in practice.

Our suggestion to phrase the quantum vacuum phenomena studied in this work in
terms of critical phenomena may be useful for future studies of possible
aspects of universality of the transition phenomenon. Specifically, second-order
phase transitions in statistical systems exhibit a large degree of universality
induced by fluctuations and near-conformal self-similar behavior in the
vicinity of the phase transition
\cite{Kadanoff:1971pc,Wilson:1973jj,Zinn-Justin:2002ecy}. The language of
critical phenomena has also been useful in the characterization of
gravitational collapse \cite{Choptuik:1992jv,Gundlach:2007gc} as well as QED
vacuum decay in terms of the Schwinger effect \cite{Gies:2015hia,Gies:2016coz}
with the degree of universality being an active research field in each of these
cases \cite{Ilderton:2015qda,Gavrilov:2019sbt,Adorno:2021xvj,Baumgarte:2023tdh,
Cors:2023ncc,Marouda:2024epb}. Within strong-field QED, a critical point has
also been discovered in the momentum spectrum for nonlinear Breit-Wheeler pair
production with the width of the photon wave packet serving as a control
parameter \cite{DegliEsposti:2023fbv}.

This article is organized as follows: \Cref{sec:simulating_vep} provides a brief
introduction to the vacuum emission picture, explains how numerical simulations
allow us to study arbitrary field configurations, and discusses the capabilities
and limitations of our current approach. \Cref{sec:collision_signatures}
elucidates the use of tailoring laser fields to achieve certain desired signal
profiles. Here, we demonstrate that our results for laser pulse collisions can
be phrased in terms of a phase transition analog. At the same time, our work
serves as an illustration of the broad abilities of the improved VacEm code.
\Cref{sec:conclusions} concludes this work and presents an outlook. The
\crefrange{sec:advancements}{sec:error_estimation} provide the technical details
for code improvements, artifact suppression and error estimation.

In line with the literature, we use Heaviside-Lorentz units with $c = \hbar =
\varepsilon_0 = 1$ and the metric $g^{\mu \nu} =
\operatorname{diag}(-1, +1, +1, +1)$ for the conceptual discussion; the numerical code as well as the description in the corresponding sections below use SI units and the metric $g^{\mu \nu} =
\operatorname{diag}(1, -1, -1, -1)$.

\section{Signal photons in the vacuum emission picture}
\label{sec:simulating_vep}

Let us briefly summarize the formalism for computing the signatures of quantum
vacuum nonlinearities in high-intensity laser pulse collisions. We pay special
attention to the assumptions, approximations and the resulting analytical error.
We focus on the numerical simulation of these signatures, discussing the
parameters which control simulation accuracy and cost.

\subsection{Vacuum emission picture} \label{sec:formalism}

We start from the one-loop Heisenberg-Euler Lagrangian to leading order in a
weak-field expansion \cite{Euler:1935zz}
\begin{equation} \label{eq:L_1-loop_weak-field}
    \begin{aligned}
        \mathcal{L}_\text{HE}^\text{1-loop}
         & = \frac{\alpha}{90 \pi}
        \frac{4 \mathcal{F}^2 + 7 \mathcal{G}^2}{E_\text{cr}^2} + O(F^6)\,,
    \end{aligned}
\end{equation}
with elementary charge $e$, electron mass $m_e$, and also setting the scale for the
critical field strength, $E_\text{cr}=\frac{m_e^2}{e}\approx \SI{1.32e18}{V/m}$.
The weak-field expansion is formulated in terms of the relativistic invariants
\begin{subequations} \label{eq:relativistic_invariants}
    \begin{alignat}{3}
        \mathcal{F} &= \frac{1}{4} F_{\mu \nu} F^{\mu \nu} 
        &&= \frac{1}{2} \left( \vect{B}^2 - \vect{E}^2 \right)\,, \\
        \mathcal{G} &= \frac{1}{4} F_{\mu \nu} \tilde{F}^{\mu \nu} 
        &&= - \vect{B} \cdot \vect{E}\,.
    \end{alignat}
\end{subequations}
Here, $F^{\mu \nu}$ denotes the field strength tensor of the applied
electromagnetic field and $\tilde{F}^{\mu \nu} = \frac{1}{2} \varepsilon^{\mu
\nu \alpha \beta} F_{\alpha \beta}$ its Hodge dual. The higher orders in
\cref{eq:L_1-loop_weak-field} are suppressed by corresponding powers of the
critical field strength, and can be derived to arbitrarily high order from the
full expression, \cf
\cite{Heisenberg:1936nmg,Schwinger:1951nm,dittrich2000,Karbstein:2019oej}.

\Cref{eq:L_1-loop_weak-field} is the relevant interaction term for the study of
signatures of quantum vacuum nonlinearities in currently achievable
high-intensity laser pulse collisions. In addition to the weak-field expansion
and the one-loop approximation, it relies on the assumption of the fields
varying slowly on scales of the (reduced) Compton wavelength
$\lambdabar_\text{C} \approx \SI{3.86e-13}{m}$ and Compton time $\tau_\text{C}
\approx \SI{1.29e-21}{s}$. For state-of-the-art and near-future laboratory
parameters, the dominant error arises from the loop expansion with the
subleading two-loop terms contributing corrections on the \SI{1}{\percent}
level. While the latter are fully computable
\cite{Ritus:1975cf,Dittrich:1985yb,dittrich2000,Gies:2016yaa}, we consider this
as an incentive to aim for a numerical error well below this higher-loop level
such that numerical predictions can reliably cover two-loop accuracy in future
studies.

In the vacuum emission picture, the signal amplitude follows as
\begin{equation} \label{eq:signal_amplitude_basic} 
    \mathcal{S}_{(p)}(\vect{k}) = \left\langle \gamma_{\beta}(\vect{k}) 
    \left| \Gamma_\text{HE}^\text{1-loop} \right| 0 \right\rangle \,,
\end{equation}
where $\langle \gamma_{\beta}(\vect{k})|$ is the one photon state with linear
polarization $\beta \in [0, 2 \pi)$ and wave vector $\vect{k}$. The interaction
of the colliding laser pulses is encoded in the effective action
$\Gamma_\text{HE}^\text{1-loop} = \int \diff{4}{x}
\mathcal{L}_\text{HE}^\text{1-loop}$, and $|0\rangle$ denotes the vacuum state.
We emphasize that \cref{eq:signal_amplitude_basic} accounts only for the
zero-to-one signal-photon transition. Processes generating two or more signal
photons are neglected, as they are typically suppressed relative to the single
signal-photon emission \cite{Karbstein:2023xmv}.

To evaluate \cref{eq:signal_amplitude_basic}, we decompose the field as
$F^{\mu \nu} \to F^{\mu \nu} + f^{\mu \nu}$ into an intense background
$F^{\mu\nu}$ treated as classical and $f^{\mu \nu}$ considered as an
operator-valued signal. For the one-signal-photon amplitude, it suffices to
consider the leading order of the expansion of \cref{eq:L_1-loop_weak-field}
around the background,
\begin{equation} \label{eq:interaction_Lagrangian}
    \begin{aligned}
        \mathcal{L}_\text{HE}^\text{1-loop}(F + f) 
        &= \mathcal{L}_\text{HE}^\text{1-loop}(F) \\
        &\quad+f^{\mu \nu} \frac{\partial\mathcal{L}_\text{HE}^\text{1-loop}(F)}
        {\partial F^{\mu \nu}} + \mathcal{O}(f^2) \,. 
    \end{aligned}
\end{equation}
Combining \crefrange{eq:L_1-loop_weak-field}{eq:interaction_Lagrangian}, we can
write the signal amplitude as \cite{Gies:2017ygp}
\begin{equation} \label{eq:signal_amplitude}
    \begin{aligned}
        \mathcal{S}_\beta (\vect{k}) = \mathcal{A} 
        & \int \diff{4}{x} \e^{\i k x} \\
        \times& \left[ 4 \left( \vect{e}_\beta \cdot \vect{E} 
        - \vect{e}_{\beta + \frac{\pi}{2}} \cdot \vect{B} \right) \mathcal{F} 
        \right.       \\
        & \left. + 7 \left( \vect{e}_\beta \cdot \vect{B} 
        + \vect{e}_{\beta + \frac{\pi}{2}} \cdot \vect{E} \right) \mathcal{G} 
        \right] \,,
    \end{aligned}
\end{equation}
where $\mathcal{A} = \frac{1}{\i} \frac{e}{4 \pi^2} \frac{m^2_e}{45}
\sqrt{\frac{\mathrm{k}}{2}} \left(\frac{e}{m^2_e}\right)^3$, and $\mathrm{k} =
k^0 = |\vect{k}|$ denotes the wave number. The linear polarization vectors
$\vect{e}_\beta(\vect{k}),\, \vect{e}_{\beta + \frac{\pi}{2}}(\vect{k})$ form an
orthonormal basis together with the normalized wave vector $\vect{e}_\vect{k} =
\vect{k}/\mathrm{k}$. We also use the short form $k x = k^\mu x_\mu$. The
(energy and emission angle resolved) differential number of signal photons is
given by
\begin{equation} \label{eq:signal_spectrum}
    \frac{\diff{3}{N_\beta (\vect{k})}}{\diff{3}{\mathrm{k}}} 
    = \frac{1}{(2 \pi)^3} \left| \mathcal{S}_\beta (\vect{k}) \right|^2 \,,
\end{equation}
where $\diff{3}{\mathrm{k}} = \mathrm{k}^2 \diff{}{\mathrm{k}} \diff{}{\Omega}$,
and $\diff{}{\Omega} = \sin(\vartheta) \diff{}{\vartheta} \diff{}{\varphi}$. In
this work, we do not resolve signal photon polarizations but sum over both
orthogonal polarization directions. Of course, the VacEm code provides fully
polarization resolved information.

\subsection{Numerical implementation} \label{sec:numerics}

The VacEm code is a numerical simulation code introduced in
\cite{Blinne:2018nbd}. Given an arbitrary EM field configuration and parameters
defining the simulation domain, it computes the signal amplitude
\cref{eq:signal_amplitude} in two parts $\mathcal{S}_\text{a/b}$ such that
$\mathcal{S}_\beta(\vect{k}) = \mathcal{A} (\sin(\beta)
\mathcal{S}_\text{a}(\vect{k}) + \cos(\beta) \mathcal{S}_\text{b} (\vect{k}))$.
In order to reduce the number of operations and enable using fast Fourier
transform (FFT) for the main computational task, the implementation works with
the reformulated equations
\begin{subequations} \label{eq:Sa_and_Sb}
    \begin{align}
        \mathcal{S}_\text{a}(\vect{k}) & = \int \diff{}{t} \e^{\i c \mathrm{k} t}
        \left[ \vect{e}_1 \cdot \vect{\hat{Q}} - \vect{e}_2 \cdot \vect{\hat{R}}
        \right] \,, \\
        \mathcal{S}_\text{b}(\vect{k}) & = \int \diff{}{t} \e^{\i c\mathrm{k} t}
        \left[ \vect{e}_2 \cdot \vect{\hat{Q}} + \vect{e}_1 \cdot \vect{\hat{R}}
        \right] \,,
    \end{align}
\end{subequations}
where $\vect{e}_1$ and $\vect{e}_2$ denote some basis vectors for the
polarization direction such that $\vect{e}_\beta(\vect{k}) = \sin(\beta)
\vect{e}_1(\vect{k}) + \cos(\beta) \vect{e}_2(\vect{k})$. We define
\begin{subequations}
    \begin{align}
        \vect{\hat{Q}} 
        & = \int \diff{3}{x} \e^{-\i \vect{k} \cdot \vect{x}} \vect{Q}
        \quad \text{with} \quad 
        \vect{Q} = 4 \vect{E} \mathcal{F} + 7 \vect{B} \mathcal{G}\,, 
        \label{eq:Q} \\
        \vect{\hat{R}} 
        & = \int \diff{3}{x} \e^{-\i \vect{k} \cdot \vect{x}} \vect{R}
        \quad \text{with} \quad 
        \vect{R} = -4 \vect{B} \mathcal{F} + 7 \vect{E} \mathcal{G}\,. 
        \label{eq:R}
    \end{align}
\end{subequations}
The numerical implementation is now straightforwardly given by $\int \diff{}{t}
\to \sum_t \Delta t$ and $\int \diff{3}{x} \e^{-\i \vect{k} \cdot \vect{x}} \to
\FFT_3$; see \cref{sec:advancements} for a pseudocode representation.

The simulation grid (in position space) is controlled by 8 parameters. There
are 4 parameters defining the simulation domain $(L_t,\, L_x,\, L_y,\, L_z)$ and
4 parameters defining the number of grid points in each dimension $(N_t,\,
N_x,\, N_y,\, N_z)$. There are multiple aspects to consider when choosing these
parameters. The simulation grid must capture the interaction of the laser
pulses in extent and resolution. Furthermore, the numerical error generally
shrinks for larger parameter values, but the computational cost grows. The
former is a minimum requirement, whereas the latter is an optimization problem.

It is useful to consider the grid resolution $(\Delta t$, $\Delta x$, $\Delta
y$, $\Delta z)$ with $\Delta \mu = L_\mu/N_\mu$. Focusing on the
counterpropagating scenario, the relevant scales for the temporal parameters are
set by the pulse durations $\tau_i$ and wavelengths $\lambda_i$, where $i \in
\mathbb{N}$ labels the laser pulse, \ie $L_t = L_t(\tau_i)$, $\Delta t = \Delta
t(\lambda_i)$. We require a sufficiently small $\Delta t$ in order to achieve
convergence of the time integration. For the spatial parameters, we connect the
longitudinal extent along the beam axis to the extent in time $L_\parallel =
L_\parallel(L_t)$ and adjust the transversal extent $L_\perp = L_\perp(w_{0,i})$
as a function of the beam waists $w_{0,i}$.

Similar considerations can be made in $\vect{k}$ space. Since $\Delta k_{x,y,z}
= 2 \pi / L_{x,y,z}$ and $k_{x,y,z,\text{max}} \approx 2 \pi / (2 \Delta
x_{x,y,z})$, the desired/required $\vect{k}$ space properties constrain the
choice of the spatial parameters $L_{x,y,z}$ and $N_{x,y,z}$.

Computational cost is only affected by $N_\mu$. The required number of
operations scales with $N_t N_x N_y N_z \log (N_x N_y N_z)$ and the required
memory with $N_x N_y N_z$. We are mainly limited by memory, in particular memory
size. \Cref{sec:advancements} provides insight into the memory usage. Memory
bandwidth and latency additionally play a crucial role when considering the
processor-memory bottleneck (performance gap) \cite{efnusheva2017}. The original
VacEm code runs on one node and calls \texttt{FFTW}
\cite{fftw2024,Frigo:2005zln} for parallel $\FFT_3$. In this work, improvements
to the computation time and memory usage were implemented. Multinode
parallelism of the time integration and single-precision float operations were
added to the feature set; see \cref{sec:advancements} and \cite{maiwald2023}. To
resolve the still existent memory bottleneck, $\FFT_3$ with distributed-memory
parallelism should be implemented in the future.

Given the current code status and hardware performance, a careful choice of all
8 simulation domain parameters can largely suppress numerical artifacts at a
reasonable computational cost. Aliasing can be avoided by choosing $\Delta \mu$
such that the sampling rate is above the Nyquist rate of the involved
frequencies. Spectral leakage caused by windowing is unavoidable and can only be
suppressed by increasing $L_{x,y,z}$. In \cref{sec:collision_signatures}, we
quantify the impact of numerical artifacts. This work demonstrates the current
capabilities of the VacEm code.

\section{Quantum-vacuum signatures from tailored-pulse collisions}
\label{sec:collision_signatures}

\subsection{Pulse collisions in a dark-field scheme} \label{sec:dark_field}

Collisions of ultra-intense laser pulses lead to the generation of scattered
photons as a signature of the fluctuation-induced effective nonlinear
interactions between electromagnetic fields  and
thus of the violation of the superposition principle by the quantum vacuum. With
two generic laser pulses, the largest signal is produced in a counterpropagating
head-on collision. However, a straightforward detection in experiment is
hampered by the fact that the quantum-induced signal photons are predominantly
scattered into the forward direction \cite{King:2015tba,Karbstein:2019oej}. This
makes it challenging to separate the signal from the huge photon background of
the driving laser pulses.

\begin{figure}
    \includegraphics[width=\columnwidth]{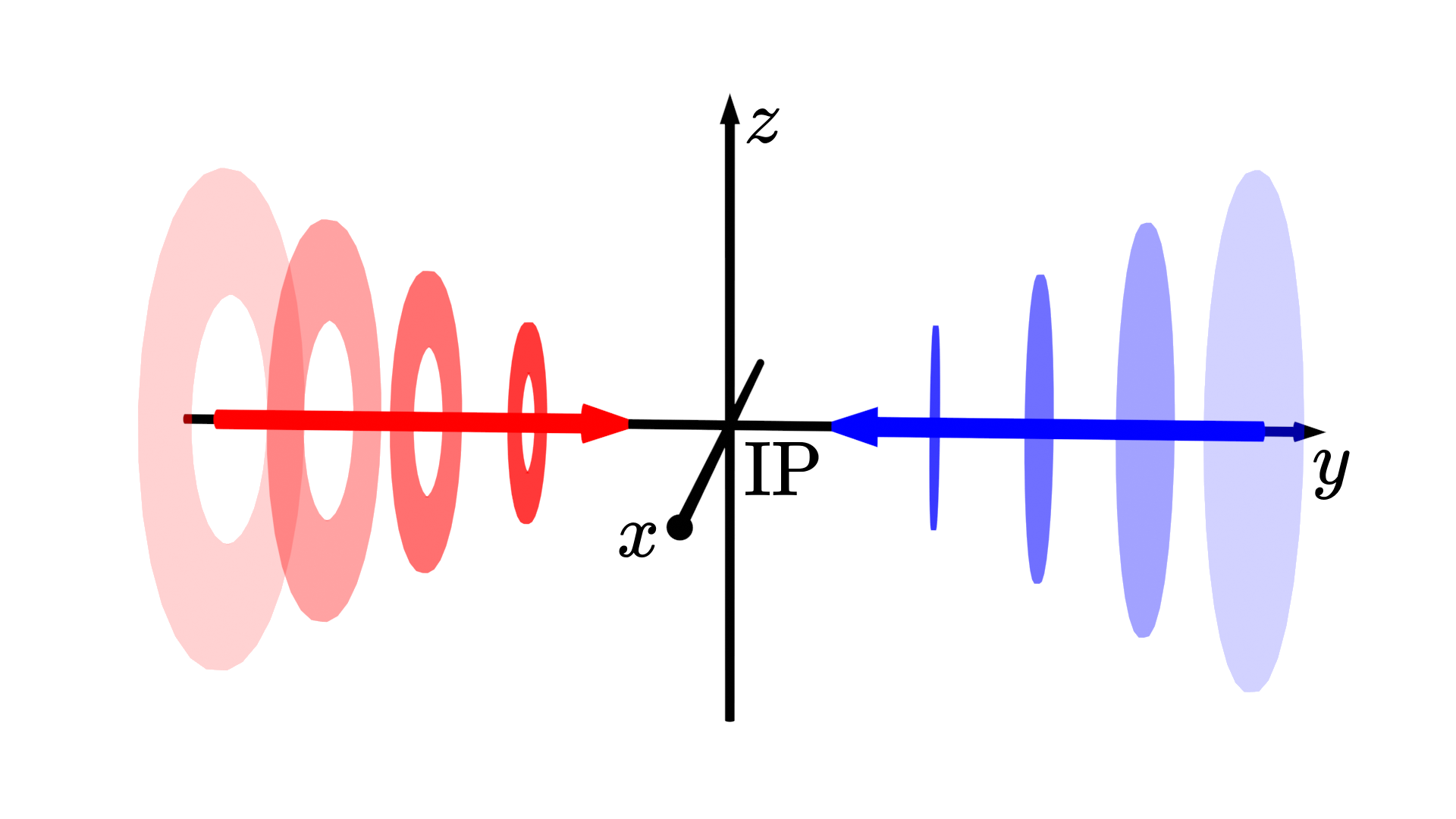}
    \caption{Sketch of the collision setup showing the annular flat-top probe in
    red propagating towards the positive $y$ direction and the
    counterpropagating Gaussian or flat-top pump in blue colliding in the
    interaction point (IP) at $(x,y,z) = (0,0,0)$. The increasing opacity
    indicates the evolution forward in time. The colored arrows indicate their
    respective propagation directions.}
    \label{fig:sketch}
\end{figure}

A promising idea to address this challenge is given by the dark-field scheme
\cite{Karbstein:2020gzg,Karbstein:2022uwf} which we also use in the following.
Here, an annular pulse is tailored by blocking a central part of the transverse
far-field profile of the incoming pulse such that the outgoing pulse also
features a field-free central region around its beam axis while a peaked on-axis
field is retained in its focus. This tailored pulse is considered as the probe
pulse going into positive $y$ direction in our setup; see \cref{fig:sketch}.
The probe pulse hits a counterpropagating pump pulse, both considered to be
optimally aligned at the collision point, which we choose as the origin of our
coordinate system. A crucial point now is that the signal photons being
scattered into the forward direction of the probe pulse have the chance to
propagate into the forward shadow of the annular pulse. If the shadow is
sufficiently dark, a detector positioned inside the dark-field region has the
chance to detect the signal photons above the shadow-suppressed background.

Focusing on the signal in the forward direction of the probe, the dominant
contribution to the signal amplitude stems from the interaction in the focus
region. Due to energy and momentum conservation, the dominant term in the signal
amplitude \cref{eq:signal_amplitude}  is linear in the probe and quadratic in
the pump pulse, \cf \cite{Karbstein:2023xmv},
\begin{equation} \label{eq:dominant_signal_contribution}
    \begin{split}
        \mathcal{S}_\beta (\vect{k})\big|_{\sim E_\text{probe}}
        \propto \int \diff{4}{x} \e^{\i k x} 
        \Big[&E_\text{probe}(x) |E_\text{pump}(x)|^2 \\
        &\times \Theta(\theta, \varphi, \beta) \Big],
    \end{split}
\end{equation}
where $\Theta$ is some function defining the details of the  angular dependence.
Therefore, it is instructive to take a closer look at the probe field and the
pump intensity in the focus.

More specifically, we use a pulsed annular flat-top beam as the probe field,
while we study both a pulsed Gaussian beam or a pulsed flat-top beam as the pump
field. In the focal $(x,z)$ plane, we choose the polarization of both beams in
the $z$ direction.

The annular flat top and its special case, the flat top, are analytically known
in the interaction region. A Maxwell solver makes them available at any time
step $t$; see \cref{sec:advancements}. For instance, the electric field
component of a linearly polarized beam (oscillation frequency $\omega_{\rm I}$)
propagating in the $y$ direction ($\rho=\sqrt{x^2+z^2}$) in the focus region is
given by
\cite{Karbstein:2023xmv}
\begin{equation} \label{eq:aft}
    \begin{aligned}
        E_\text{I}(t,\vect{x}) 
        = 2& E_0 \sqrt{\frac{1 - 1/\e}{1+\nu}} 
        \,\e^{- \left( \frac{y - t}{\tau_\text{I}/2} \right)^2} \\
        & \times \left[ \frac{J_1\!\left( \sqrt{
        \frac{2(1 - 1/\e)}{1 + \nu}}
        \frac{2 \rho}{w_\text{I}} \right)}{\sqrt{
        \frac{2(1 - 1/\e)}{1 + \nu}} 
        \frac{2 \rho}{w_\text{I}}} \right. \\
        & \qquad \left. - \nu 
        \frac{J_1\!\left( \sqrt{
        \frac{2(1 - 1/\e)}{1 + \nu}}
        \nu 
        \frac{2 \rho}{w_\text{I}} \right)}{\sqrt{
        \frac{2(1 - 1/\e)}{1 + \nu}}
        \nu
        \frac{2 \rho}{w_\text{I}}} \right] \\
        & \times \cos\left( \omega_\text{I} (y - t) \right) \,,
    \end{aligned}
\end{equation}
with the corresponding Gaussian peak field amplitude \cite{Karbstein:2019bhp}
\begin{equation}
    E_0 = \sqrt{8 \sqrt{\frac{2}{\pi}} 
    \frac{W_\text{I}}{\pi w_\text{I}^2 \tau_\text{I}}} \,,
\end{equation}
expressed in terms of the pulse energy $W_{\rm I}$, waist $w_{\rm I}$, and $1/{\rm e}^2$ (on intensity level) pulse duration $\tau_{\rm I}$. The subscript $\mathrm{I}$ emphasizes the validity limited to the
interaction region, $J_1$ is the Bessel function of first kind at order $1$, and
$\nu=(\theta_{\rm in}/\theta_{\rm out})^2$ is the blocking fraction for the
annular flat top with inner and outer far-field radial divergences
$\theta_\text{in}$, $\theta_\text{out}$. In the remainder of this work, we assume
$\nu = 1/4$ for the annular flat top and, of course, $\nu = 0$ for the flat top.

\begin{figure}
    \includegraphics[width=\columnwidth]{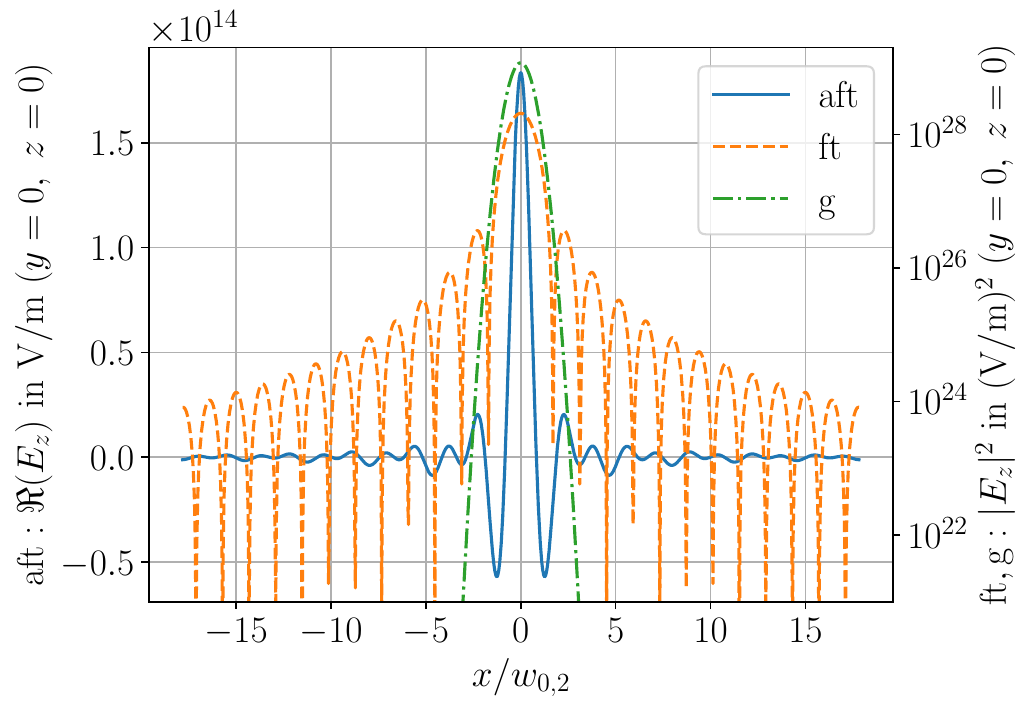}
    \caption{Transverse profile of the probe amplitude (aft) and the pump
    intensity (ft, g) in the focus; \cf \cref{eq:dominant_signal_contribution}
    (aft: annular flat top; ft: flat top; g: Gaussian; $w_{0,2}/w_{0,1} = 1.9$;
    see text for all other relevant parameters).}
    \label{fig:transverse_pulse_profiles}
\end{figure}

In \cref{fig:transverse_pulse_profiles}, we depict the transverse focus profiles
of the annular flat-top beam amplitude, and of the intensities for the flat top
beam and the Gaussian beam along the transversal $x$ direction; \cf
\cref{eq:dominant_signal_contribution}. Throughout this work, the pulse
parameters are given by the wavelength $\lambda = \SI{800}{nm}$, pulse duration
$\tau_\text{FWHM} = \SI{20}{fs}\ (= \sqrt{\ln(2)/2}\ \tau)$, and pulse energy $W
= \SI{25}{J}$ corresponding to a petawatt-class high-intensity pulse. We
somewhat arbitrarily keep the probe's beam waist fixed at $w_{0,1} \approx
\SI{2.18}{\micro m}$. The pump's beam waist $w_{0,2}$ is varied to obtain
different ratios $w_{0,2}/w_{0,1}$. While the relative polarization could be
optimized for certain observables such as polarization flip or total photon
yield \cite{Karbstein:2019bhp}, it is not really of importance for our purpose
since it just modifies the overall factor for the signal amplitude.

From \cref{fig:transverse_pulse_profiles}, it is obvious that the Gaussian beam
has a Gaussian transversal profile in the focal plane. By contrast, the flat-top
beams exhibit a nontrivial Airy-ring structure in the focal plane. Because the
quantum vacuum signal is peak-field driven, it is clear that the dominant
contribution to the signal will arise from the focal region where the maxima of
both pulses have the largest overlap. On the other hand, it is also clear that
the precise angular distribution of the signal photons depends on the details of
the laser beam structure. In particular, the much richer transverse structure of
a flat-top pump suggests a qualitative difference of the behavior of the angular
signal distribution in comparison to that driven by a Gaussian pump.

\subsection{A phase transition analog} \label{sec:pt_analogue}

In the present work, we focus on a specific phenomenon of the angular
signal-photon distribution which can be phrased in the language of a phase
transition. As symmetries play a prominent role in the theory of phase
transitions, we first note that in the present counterpropagating setup the
composite field profile of the colliding laser pulses features a rotational O(2)
symmetry about the propagation axis. This rotational symmetry is broken on the
level of the field vectors. It turns out that this breaking, of course, results
in a similar breaking of the rotational symmetry of the quantum-induced signal
photons in terms of their transversal polarization state. In addition, the number of
signal photons depends on the relative polarization of the colliding pulses
\cite{Karbstein:2019bhp}. Nevertheless, the polarization-averaged signal-photon
amplitude again inherits the rotationally O(2) symmetry.

The following argument illustrates that such a phase transition analog with
respect to the realization of this symmetry must exist. In view of
\cref{fig:transverse_pulse_profiles}, we can envision the possible interaction
scenarios depending on the ratio of pump beam waist $w_{0,2}$ and the probe beam
waist  $w_{0,1}$. Regardless of the pump type, there exist two limiting cases
\cite{Karbstein:2020gzg}: In the limit of a small pump beam waist
$w_{0,2}/w_{0,1} \ll 1$, the pump predominantly interacts with the main peak of
the annular flat top, which effectively corresponds to the collision of two
pulses with a single-peak focus structure, similar to two Gaussian peaks. Hence,
we expect the signal to also exhibit a single peak structure which---due to the
O(2) symmetry---should lie on axis in the forward direction (positive $y$ axis).

In the opposite limit of a large pump beam waist $w_{0,2}/w_{0,1} \gg 1$, the
pump pulse can be well approximated by a plane wave over the extent of the probe
focus region. This implies that the pump  cannot transfer a relevant amount of
transverse momentum to the probe field.  We thus expect the angular distribution
of the signal to resemble the probe in the far field. For the present case of an
annular probe pulse, we also expect the signal  to
feature an annular structure in the transverse $(x,z)$ plane.

For instance, for the signal intensity along the $x$ axis at $z=0$ and in the far
field at sufficiently large $y$, we expect a single-peak structure for
$w_{0,2}/w_{0,1} \ll 1$ with the peak at $x=0$ on axis, and a double-peak
structure for  $w_{0,2}/w_{0,1} \gg 1$ with the peaks at $|x|>0$ off axis.
(Below, we use an angle variable instead of the $x$ axis, but the
features remain the same.) Flipping the sign of the signal-photon distribution
in the far field, its shape and behavior resemble that of a
Landau-Ginzburg-type potential for an order parameter across a phase transition.

For intermediate values of $w_{0,2}/w_{0,1}$, it is clear now that a transition
between a single- and a double-peak structure needs to take place. From the
analog to a phase transition, we can infer that this transition can occur in
various ways: The transition can be smooth, similar to a second-order transition
such that the positions of the double peaks evolve continuously from the single
peak for increasing $w_{0,2}/w_{0,1}$ as found analytically for the collision of
an annular flattened-Gaussian probe with a Gaussian pump beam in
\cite{Karbstein:2020gzg}. Alternatively, the double peaks can appear in addition
to the single peak exhibiting a coexistence of all peaks for a certain interval
of $w_{0,2}/w_{0,1}$ values. This would correspond to a first-order phase
transition.

\subsection{Numerical simulation} \label{sec:numerical_simulation}

Using the VacEm code with improved performance, we show in the following that
both types of transitions can occur depending on the properties of the pump
(flat-top \vs Gaussian pump). This difference can again be motivated from
\cref{fig:transverse_pulse_profiles}. For the case of a Gaussian pump,
continuously more and more side peaks (Airy rings) in the focal region of the
annular flat top contribute to the interaction the larger the ratio
$w_{0,2}/w_{0,1}$. By contrast, for the flat-top pump, the side peaks on the
fringes of the focal region go through cycles of alignment and misalignment when
varying $w_{0,2}/w_{0,1}$.

Our investigation of the two setups of counterpropagating axisymmetric pulse
collisions defined above uses the following simulation parameters: $L_t = 4
\tau,\ L_y \approx 6 c \tau,\ L_{x,z} \approx 18.5 w_{0,2,\text{max}}$ and a
grid resolution $\Delta \mu$ equivalent to 7.5 points per period $\lambda/c$ and
wavelength $\tilde{k}_{x,y,z}/(2 \pi)$, where $\tilde{k}_{x,y,z}$ are the
estimated maximum wave numbers along each spatial axis. At this resolution, we
resolve frequencies up to $(\lambda/3)^{-1}$ along the propagation direction.
This would even be sufficient to account for photon merging
\cite{DiPiazza:2005jc,Fedotov:2006ii,Gies:2014jia,Bohl:2015uba,Gies:2016czm,Sasorov:2021anc,Sundqvist:2023hvw},
a nonlinear phenomenon that is found to be strongly suppressed in our setup.

The reasoning behind the choice of $L_{x,y,z}$ is discussed in
\cref{sec:artifacts}. The simulations make use of single-precision
floating-point operations. The latter provide a significant increase in
efficiency at basically no impact on the final accuracy of our result, as the
numerical error is dominated by other sources; see \cref{sec:advancements}. The
pump's beam waist $w_{0,2}$ is varied between \SI{2.0}{\micro m} and
\SI{8.0}{\micro m} in steps of \SI{0.67}{\micro m}, \ie $w_{0,2}/w_{0,1}$
between \num{0.92} and \num{3.7} in steps of \num{0.31}. This gives us 10 values
for $w_{0,2}$ and encompasses the expected transition region. After establishing
this broad range, we zoom in to better resolve the transition. An additional 10
data points are obtained for $w_{0,2}$ between \SI{4.1}{\micro m} and
\SI{4.6}{\micro m} in steps of \SI{0.061}{\micro m}, \ie $w_{0,2}/w_{0,1}$
between \num{1.9} and \num{2.1} in steps of \num{0.028} in order to study the
nature of the transition.

\begin{figure}
    \includegraphics[width=\columnwidth]{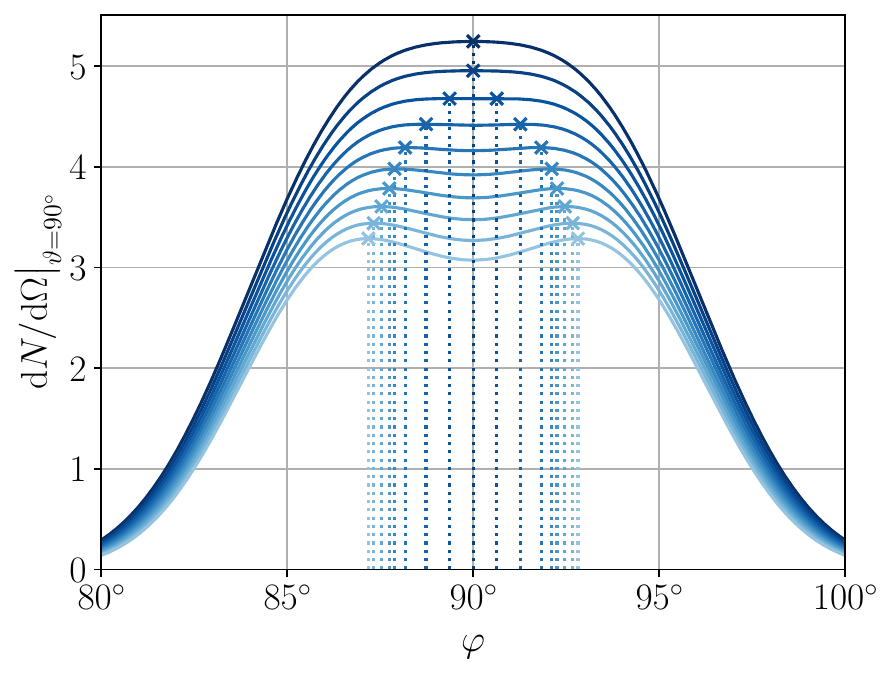}
    \caption{ Differential number of
        signal photons in the $(x,y)$ plane as a function of the azimuthal
        angle with $\varphi=\SI{90}{\degree}$
        denoting the forward propagation direction of the probe (annular flat top)
        and signal. For the Gaussian pump setup, we observe the continuous
        emergence of a double-peak structure with the pump waist $w_{0,2}$
        between \SI{4.1}{\micro m} (dark blue) and \SI{4.6}{\micro m} (light
        blue) in steps of \SI{0.061}{\micro m}. The peaks of the individual
        curves are marked by a cross $(\times)$. This transition is analogous to
        a second-order phase transition.}
    \label{fig:vbs_xz_9_anim_comp}
\end{figure}

\begin{figure}
    \includegraphics[width=\columnwidth]{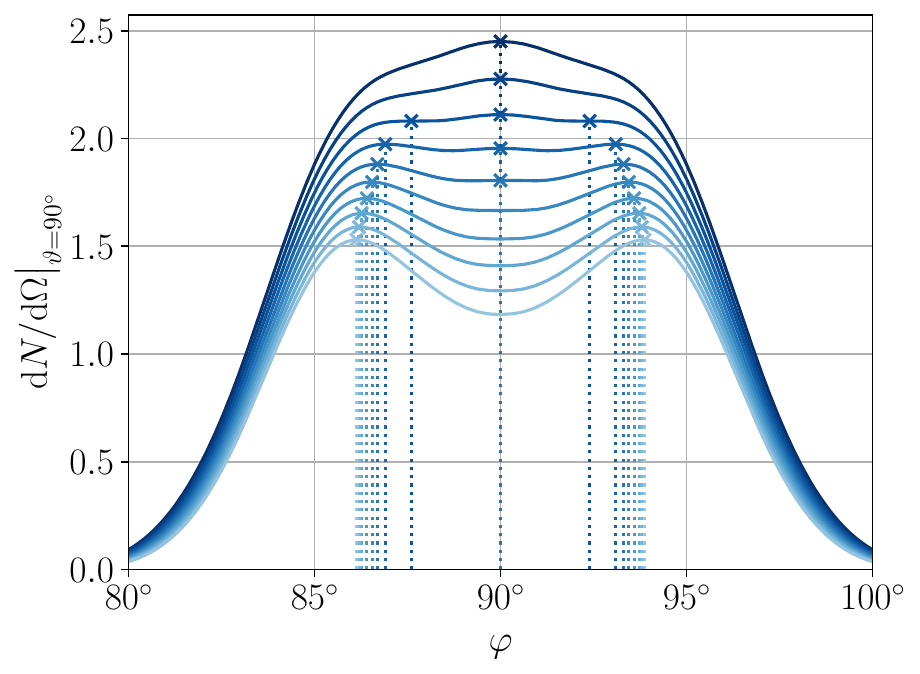}
    \caption{Differential number of
    signal photons in the $(x,y)$ plane as a function of the azimuthal angle
    with $\varphi=\SI{90}{\degree}$ denoting the
    forward propagation direction of the probe (annular flat top) and signal. For
    the flat-top pump setup, the double-peak structure emerges while the central
    peak still exists. The graphs span a range of pump waists $w_{0,2}$ between
    \SI{4.1}{\micro m} (dark blue) and \SI{4.6}{\micro m} (light blue) in steps
    of \SI{0.061}{\micro m}. The peaks of the individual curves are marked by a
    cross $(\times)$. This transition is analogous to a first-order phase
    transition.}
    \label{fig:vbs_flat_xz_9_anim_comp}
\end{figure}

We employ the VacEm code to simulate the signal amplitude
\cref{eq:signal_amplitude} for each value of $w_{0,2}$. From this, we compute
the corresponding differential number of signal photons
\cref{eq:signal_spectrum}. Our observable of choice for visualizing the
resulting data is $\mathrm{d} N / \mathrm{d} \Omega \big|_{\vartheta =
\SI{90}{\degree}}$; \ie we integrate the differential number of signal photons
over $\mathrm{k}$ and study it in the $(x,y)$ plane at polar angle
$\vartheta=\SI{90}{\degree}$. The spherical coordinate system
$(r,\vartheta,\varphi)$ is defined such that $r$ is the radial distance,
$\vartheta$ is the polar angle between the $z$ axis and the radial axis, and
$\varphi$ is the azimuthal angle between the $x$ axis and the radial axis.
\Cref{fig:vbs_xz_9_anim_comp,fig:vbs_flat_xz_9_anim_comp} show the signal
profiles for the range of pump beam waists $w_{0,2}$ in the vicinity of the
phase transition as a function of the azimuthal angle $\varphi$ near and across
the forward propagation direction $\varphi=\SI{90}{\degree}$.

In \cref{fig:vbs_xz_9_anim_comp}, we depict the data for the collision of the
annular flat-top probe with the Gaussian pump for pump waists $w_{0,2}$ between
\SI{4.1}{\micro m} (dark blue) and \SI{4.6}{\micro m} (light blue) in steps of
\SI{0.061}{\micro m}. The topmost dark blue curve in
\cref{fig:vbs_xz_9_anim_comp} belongs to the smallest pump waist $w_{0,2}$,
exhibits a single-peak structure on the propagation axis at
$\varphi=\SI{90}{\degree}$, and resembles a flattened Gaussian \cite{gori1994}.
For even smaller $w_{0,2}$ (not shown in the figure), the signal profile
converges towards a Gaussian shape. For increasing pump waist $w_{0,2}$, the
central peak broadens even further and evolves smoothly into a double peak with
the two peaks moving further outward away from the propagation axis
\cite{Karbstein:2020gzg}. The main emission directions are marked by a cross
$(\times)$. The transition from the single peak on axis to the double peaks off
axis is continuous and---using the peak position $\varphi_\text{peak}$ (in the
branch $\varphi \geq \SI{90}{\degree}$) as an order parameter---resembles a
phase transition of second order.

In \cref{fig:vbs_flat_xz_9_anim_comp}, we show the data for the collision of the
same probe (annular flat top) with the flat-top pump for the same range of pump
beam waists.  The particular features of the topmost dark blue curve can be
explained by the already growing left and right peaks. For smaller $w_{0,2}$ (not
shown in the figure), this effect is not yet visible, and a Gaussian shape is
approached. For an increasing pump beam waist $w_{0,2}$, the side peaks become more
pronounced, and we observe a coexistence region of $w_{0,2}$ values where the
side peaks as well as the central peak are local maxima marked by a cross
$(\times)$. This coexistence makes the transition resemble a first-order phase
transition. For larger $w_{0,2}$ beyond the coexistence region, the
differential number of signal photons on the
propagation axis at $\varphi=\SI{90}{\degree}$ becomes a local minimum. The
lowermost light blue curve belongs to the largest $w_{0,2}$ values of the plot.

From the markings, a qualitative difference in the transition process, as
compared to the analogous scenario with a fundamental Gaussian pump, is evident.
Counterpropagating axisymmetric collision of the annular flat-top probe and the
Gaussian pump  in \cref{fig:vbs_xz_9_anim_comp} shows a continuous formation of
the left and right peaks from the center. The central peak effectively dissolves
into the outer peaks. In contrast, the counterpropagating axisymmetric collision
of the annular flat-top probe  and the flat-top pump  in
\cref{fig:vbs_flat_xz_9_anim_comp} shows that the outer peaks start to form
independently of the central peak. There is even a period during the transition
where the outer peaks and the central peak exist simultaneously.

Both figures clearly reflect the expectations of the limiting cases: For a small
pump beam waist, only the interactions of the central focal peaks are relevant
and lead to a formation of a signal peak on axis. For larger pump beam waists,
the pump eventually resembles a plane wave and the signal takes over the annular
far-field profile of the probe. Of course, for the beam waists shown in the
figures, we are still far away from the plane-wave limit $w_{0,2}/w_{0,1}\to
\infty$.

\begin{figure}
    \includegraphics[width=\columnwidth]{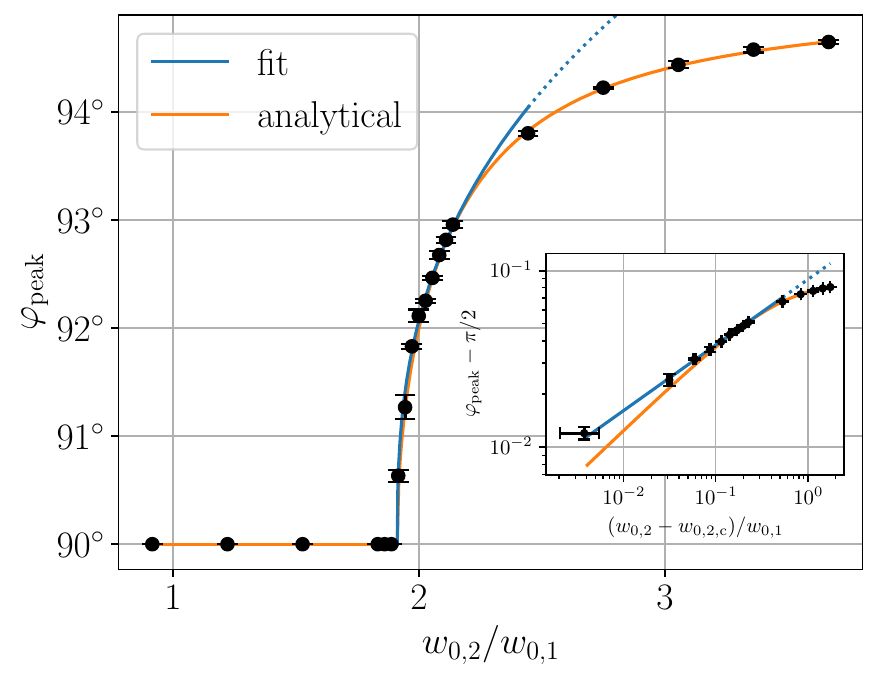}
    \caption{Main emission direction of the far-field signal-photon distribution
    for an annular flat-top probe colliding with a Gaussian pump as a function
    of the pump-to-probe waist ratio $w_{0,2}/w_{0,1}$ across the transition.
    Interpreting the main emission direction (in the branch $\varphi \geq
    \SI{90}{\degree}$) as an order parameter, we observe a continuous
    second-order phase transition. The blue curve represents the power-law fit
    \cref{eq:critical_param_fit} for the critical parameters of the transition
    region. The orange curve depicts the analytical
    estimate~\eqref{eq:phipeak_analyt}. The double logarithmic plot in the inset
    illustrates that the critical region is well described by this simple
    scaling law \cref{eq:critical_param_fit} with a critical exponent $\boldsymbol{\beta}$.}
    \label{fig:vbs_xz_4_5_meta_phi_peak_result}
\end{figure}

\begin{figure}
    \includegraphics[width=\columnwidth]{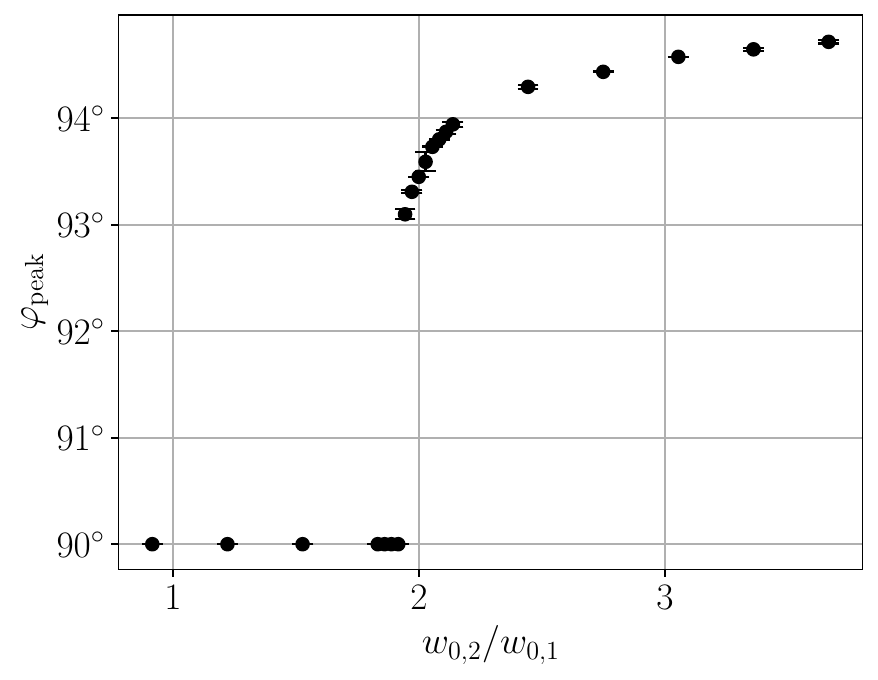}
    \caption{Main emission direction of the far-field signal-photon distribution
    for an annular flat-top probe colliding with a flat-top pump as a function
    of the pump-to-probe waist ratio $w_{0,2}/w_{0,1}$ across the transition.
    Interpreting the main emission direction (in the branch $\varphi \geq
    \SI{90}{\degree}$) as an order parameter, we observe a discontinuous
    first-order phase transition marked by a finite jump of the order
    parameter.}
    \label{fig:vbs_flat_xz_1_2_meta_phi_peak_result}
\end{figure}

To further quantify the transition, we study the dependence of the main emission
direction of the signal photons $\varphi_\text{peak}$ (in the branch $\varphi
\geq \SI{90}{\degree}$) on the beam-waist ratio $w_{0,2}/w_{0,1}$ in
\cref{fig:vbs_xz_4_5_meta_phi_peak_result,fig:vbs_flat_xz_1_2_meta_phi_peak_result}.

In these plots, we include a broader range of $w_{0,2}$ values than used in
\cref{fig:vbs_xz_9_anim_comp,fig:vbs_flat_xz_9_anim_comp}. In the language of
critical phenomena, it is obvious from these figures that the beam-waist ratio
$w_{0,2}/w_{0,1}$ can be identified as a control parameter while the peak
position $\varphi_\text{peak}$ serves as an order parameter.

For the case of a Gaussian pump (\cref{fig:vbs_xz_4_5_meta_phi_peak_result}),
the second-order nature of the phase transition is not only confirmed by the
qualitative aspects of the graph. We can also quantify the analogy further by
fitting the data to a power law as suggested by conventional critical phenomena.
Using the fit model
\begin{equation} \label{eq:critical_param_fit}
    \varphi_\text{peak}\big|_\text{transition}
    = C \left(\frac{w_{0,2} - w_{0,2,\text{c}}}{w_{0,1}}\right)^{\boldsymbol{\beta}}
    + \frac{\pi}{2},
\end{equation}
with parameters $C$, $\boldsymbol{\beta}$, and $w_{0,2,\text{c}}$, we find for
the critical exponent of the order parameter $\boldsymbol{\beta} = \num{0.372
\pm 0.014}$. For the critical beam waist, we obtain
$w_{0,2,\text{c}}\big|_\text{g} = \SI{4.1736 \pm 0.0037}{\micro m}$, implying a
precise result for the critical beam-waist ratio
$(w_{0,2,\text{c}}/w_{0,1})\big|_\text{g} = \num{1.9121 \pm 0.0017}$ which plays
the role of the critical point. The fit includes all data points of
\cref{fig:vbs_xz_4_5_meta_phi_peak_result} with $ \varphi_\text{peak} >
\SI{90}{\degree}$ except for the five right-most points. The errors account for
only a part of the discretization errors; see discussion below. In the inset of
\cref{fig:vbs_xz_4_5_meta_phi_peak_result}, the double logarithmic
representation demonstrates that the power law \cref{eq:critical_param_fit}
indeed describes the critical region rather accurately over 2 orders of
magnitude.

An analytical estimate for the main emission direction for the scenario with a
Gaussian pump can be readily extracted from Eq.~(3.110) of
\cite{Karbstein:2023xmv}. Demanding the second derivative of this expression for
$\vartheta$ to vanish identically, we infer
\begin{equation} \label{eq:phipeak_analyt}
 \varphi_\text{peak} = \delta\varphi + \frac{\pi}{2}\,, 
\end{equation}
with $\delta\varphi=0$ for $w_{0,2}\leq w_{0,2,\text{c}}$ and $\delta\varphi$
implicitly defined by
\begin{equation}
        I_1\bigl(A\tfrac{w_{0,2}}{\lambda} \delta\varphi\bigr)\,
        \mathrm{e}^{-\frac{1-\nu}{2}(\frac{A}{\pi})^2}
        =\sqrt{\nu}\,
        I_1\bigl(\sqrt{\nu}\,A\tfrac{w_{0,2}}{\lambda} \delta\varphi\bigr)
        \label{eq:BesselI1}
\end{equation}
for  $w_{0,2}\geq w_{0,2,\text{c}}$. Here, $I_1$ denotes the modified
Bessel function of the first kind and $A=\pi\sqrt{\frac{2(1-1/{\rm
e})}{1+\nu}}\frac{w_{0,2}}{w_{0,1}}$.

The analytical estimate for the critical beam-waist ratio is
\cite{Karbstein:2023xmv}
\begin{equation}
    (w_{0,2,\text{c}}/w_{0,1})\big|_\text{g}
    =\sqrt{\frac{1}{1-1/\e}\frac{1+\nu}{1-\nu}\ln(1/\nu)}\,,
    \label{eq:CPanalytical}
\end{equation}
which yields $(w_{0,2,\text{c}}/w_{0,1})\big|_\text{g}\simeq1.9118$ for
$\nu=1/4$ as considered in the present work. Applying the fit model
\cref{eq:critical_param_fit} to the analytical estimate \cref{eq:phipeak_analyt}
in the same range (but starting at the critical point) yields $\boldsymbol{\beta} = \num{0.4370 \pm 0.0038}$. For the
critical point, we observe a remarkable agreement of the analytical estimate
resorting to an {\it infinite Rayleigh range approximation} with the result of
the full numerical calculation. The critical exponent shows a larger deviation.
In fact, it is expected that the critical point can be more accurately fitted
than the critical exponent since the number and range of data points is small
and therefore sufficiently captures only the \textquote{location} as a local
property but not the \textquote{steepness} as an extended feature of the
transition. Additionally, it is important to consider that the error in
\cref{fig:vbs_xz_4_5_meta_phi_peak_result} is significantly smaller than the
total numerical error; see discussion below.

We note that the critical point could, of course, be straightforwardly inferred
also from the discretized scan of the parameter space. At the current
discretization step size, this yields $(w_{0,2,\text{c}}/w_{0,1})\big|_\text{g}
= \num{1.902 \pm 0.014}$ which is consistent with the fit result but less
precise.

For the case of a flat-top pump the order parameter undergoes a first-order
phase transition analog as shown in
\cref{fig:vbs_flat_xz_1_2_meta_phi_peak_result}. The jump in the order parameter
goes hand in hand with the observation of a coexistence region as it is obvious
from \cref{fig:vbs_flat_xz_9_anim_comp}. The critical point in terms of the
critical beam-waist ratio for the flat-top pump setup is given by
$(w_{0,2,\text{c}}/w_{0,1}) \big|_\text{ft} = \num{1.930 \pm 0.014}$ based on
the discretized scan in $w_{0,2}$.

Note that the critical points for the Gaussian pump and the flat-top pump
setup are close. Because the flat-top pump considered here can be understood as
the limiting case of an infinite-order flattened-Gaussian beam
\cite{gori1994,Karbstein:2023xmv}, the class of flattened-Gaussian beams
naturally provides a smooth interpolation between Gaussian and flat-top beams.
Therefore, we expect a continuous interpolation between the phase transition
analogs in
\cref{fig:vbs_xz_4_5_meta_phi_peak_result,fig:vbs_flat_xz_1_2_meta_phi_peak_result}.
Since the flattened-Gaussian beams of order one or higher exhibit an Airy-ring
structure, it remains an interesting open question for the future as to whether
the transition from second to first order occurs at a critical value of the
order of the flattened Gaussian.

As discussed in much more detail in \cref{sec:numerics}, the computations
leading to the results presented in this section are subject to numerical errors
and artifacts. With regard to the transverse focus profiles involved, \cf
\cref{fig:transverse_pulse_profiles}, it is clear that resolving the flat-top
beams requires particular care. Since the simulation is confined to a finite
volume, we unavoidably lose information. Specifically, cutting off the spatial
directions transverse to the beam axes of the driving laser fields goes along
with a loss of information that is significant for preserving the relevant
beam-profile information.

In the focus, the flat-top structure is encoded in the transversal structure on
the fringes of the central peak. Therefore, it is advisable to choose the
transversal domain such that the field is cut off at extrema in the focal plane.
In this way, the spectral leakage during the $\FFT_3$ caused by the
discontinuities in the periodic continuation is minimized. For the estimate of
the error $\Delta \varphi_\text{peak}$ in
\cref{fig:vbs_xz_4_5_meta_phi_peak_result,fig:vbs_flat_xz_1_2_meta_phi_peak_result},
we have repeated the simulation for a range of values of the transversal length
parameters $2 w_{0,2,\text{max}} \leq L_{x,z} \lessapprox 18.5
w_{0,2,\text{max}}$. A Richardson extrapolation \cite{richardson1911} is then
used to obtain an error estimate. Thus, the error corresponds to the limit of
accounting for the complete Airy-ring structure of the annular flat-top probe
and its effect on the signal's main emission direction; \ie the qualitative
features of the phase transition analogs. A detailed discussion of the
variation of $L_{x,z}$ in order to estimate the resulting error is provided in
\cref{sec:error_estimation}.

Additionally, we discuss the impact of the finite propagation lengths $L_{t,y}$
and grid resolution $\Delta \mu$. The estimated errors from these
discretization parameters for the qualitative features of
\cref{fig:vbs_xz_4_5_meta_phi_peak_result,fig:vbs_flat_xz_1_2_meta_phi_peak_result}
are found to be small. An estimate for the total discretization error can be
obtained by assuming linear convergence in all discretization parameters. In
this case, we obtain $(w_{0,2,\text{c}}/w_{0,1})\big|_\text{g} = \num{1.9065 \pm
0.0066}$ and $\boldsymbol{\beta} = \num{0.417 \pm 0.083 }$, which is in
satisfactory agreement with both the simulation result elaborated above and the
analytical estimate. Details are provided in \cref{sec:error_estimation}.

\subsection{Universality}

For the laser pulse collisions studied above, the language of critical phenomena
is  useful to qualitatively classify the signal emission phenomena as well as to
quantify the second-order transition in terms of a critical scaling law and a
corresponding critical exponent for the order parameter. This raises the
interesting question to which extent the concept of universality applies to the
presently studied observables. 

For the preceding examples, our results indicate that the qualitative difference
between the first- and second-order transitions originates in the transverse
focus profile. Increasing the waist of the Gaussian pump, the overlap with the
Airy rings of the probe changes smoothly, leading to a smooth transition. By
contrast, the overlap for the case of a flat-top pump and the annular probe goes
through a sequence of commensurate and incommensurate overlaps, representing a
source for a discontinuous transition. Since interpolations between Gaussian and
flat-top beam profiles exist in the form of flattened Gaussians
\cite{Karbstein:2023xmv}, it is natural to expect that the observables in the
critical region depend on the transversal shape of the pulse. For instance, the
critical exponent $\boldsymbol{\beta}$ is likely to depend on the specifics of
the beam choice. This would correspond to a lower degree of universality in
comparison to critical phenomena in statistical physics.

As an illustration, let us consider again the second-order transition for the
case of a Gaussian pump as a function of the blocking fraction $\nu$ for the
annular flat-top probe beam. It is obvious that the critical exponent
$\boldsymbol{\beta}$ cannot be independent of $\nu$: In the limit $\nu\to 0$,
the beam turns from annular to standard flat-top form and thus the phase
transition has to vanish. However, for a large blocking fraction $\nu \to 1-
\epsilon$, $0<\epsilon\ll 1$, we expect the phase transition to occur as a
stable phenomenon. In order to check for universality, we use the infinite
Rayleigh range approximation yielding \cref{eq:BesselI1,eq:CPanalytical} for an
analytical estimate of the critical point and the critical exponent as a
function of $\nu$. 

\begin{figure}
    \includegraphics[width=\columnwidth]{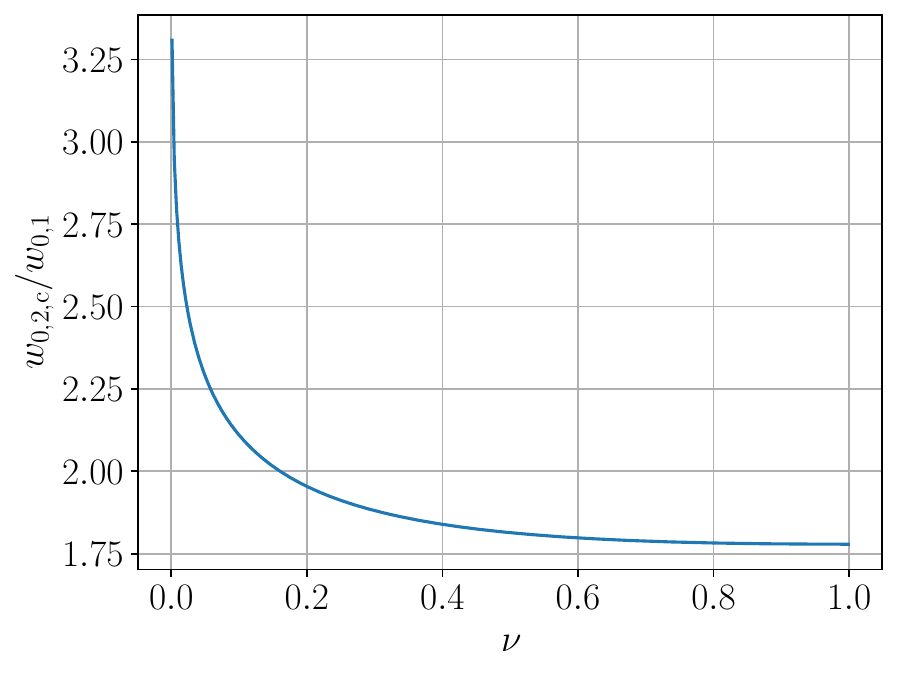}
    \caption{Analytical estimate \cref{eq:CPanalytical} for the location of the
    critical point in the beam-waist ratio (onset of a second order phase
    transition in the emission direction) as a function of the annular-beam
    blocking fraction $\nu$ for the collision with a Gaussian beam. For $\nu\to
    0$, the probe beam becomes a flat top and thus the phase transition
    disappears. For any value of $\nu\in (0,1)$, a second-order phase transition
    is present.}
    \label{fig:CPofnu}
\end{figure}

\Cref{fig:CPofnu} displays the value of the critical point in the beam waist
ratio as a function of $\nu$. We observe that the critical point indeed tends to
infinity for $\nu\to 0$, implying that the phase transition disappears as
expected for $\nu=0$. 

\begin{figure}
    \includegraphics[width=\columnwidth]{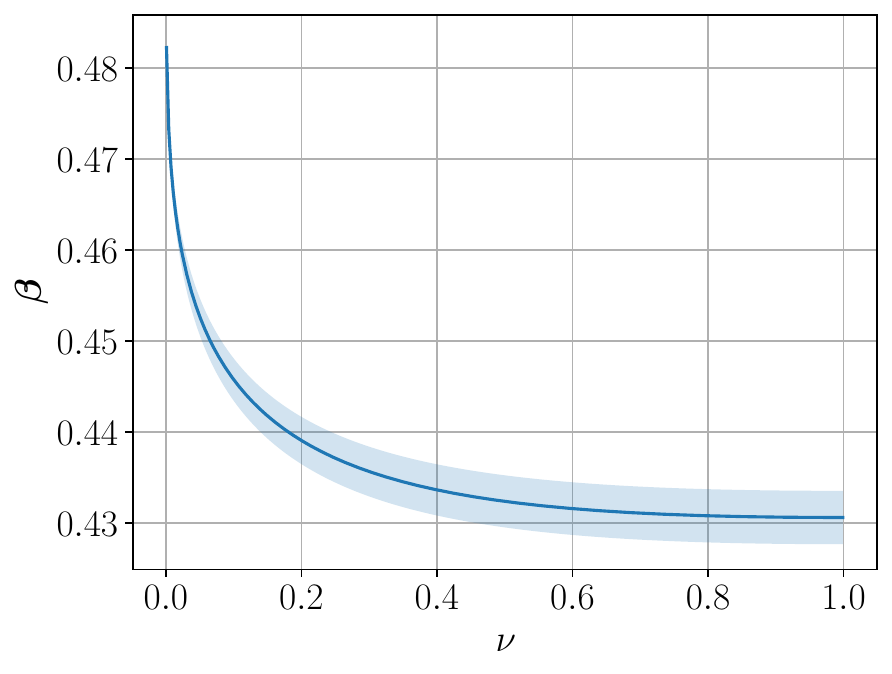}
    \caption{Analytical estimate based on \cref{eq:BesselI1} for the critical
    exponent $\boldsymbol{\beta}$ of the second order phase transition in the
    emission direction as a function of the annular-beam blocking fraction $\nu$
    for the collision with a Gaussian beam. An effective universality is
    observed for larger values of $\nu$, where the far field of the annular
    probe becomes a thin ring. The blue-shaded region represents the $1\sigma$
    error band.}
    \label{fig:betaofnu}
\end{figure}

\Cref{fig:betaofnu} depicts $\boldsymbol{\beta}(\nu)$, exhibiting a large
variation for small $\nu$ where the phase transition tends to vanish.
Interestingly, the critical exponent $\boldsymbol{\beta}(\nu)$ is rather
independent of $\nu$ for larger values of the blocking fraction where the
annular nature of the beam becomes more and more pronounced. In the idealized
limit $\nu \to 1^{-}$ where the annular beam is an infinitesimally thin ring of
light in the far field, we approximately find $\boldsymbol{\beta}\simeq 0.43$.

In summary, this simple example demonstrates that the second-order phase
transition analog does not extend to quantitative universality in the sense of
distinct universality classes characterized by specific values for the critical
exponents independent of the precise realization of the phenomenon. Still, we
find indications for an effective universality in the sense of the critical
exponent depending only weakly on realization details in certain parameter
regions.  

\section{Conclusions and outlook} \label{sec:conclusions}

We have performed a theoretical study of the quantum vacuum signal attainable in
collisions of ultra-intense laser pulses based on the effective action of QED.
As an increasing number of petawatt-class lasers are coming online nowadays,
increasingly refined theoretical methods in combination with efficient detection
schemes are necessary to exploit the full potential of these facilities for the
discovery and exploration of the nonlinear response of the quantum vacuum at
highest intensities.

In the present work, we demonstrate that the qualitative features of observables
can depend strongly on the details of the laser pulses, making the development
and use of powerful simulation tools important in the future. For the simple
case of a head-on collision of counterpropagating pulses and using a recently
proposed dark-field scheme that involves an annular probe, we discovered that
observables such as  differential numbers of signal photons can undergo the
analog of a phase transition as a function of the beam-waist ratio. We
highlighted that the nature of this phase transition can depend on the details
of the pump beam profile.

The existence of this phase transition analog can be understood from the pulse
properties in the focal region that are most relevant for the quantum signal
generation. The way that the signal realizes the axisymmetry of the setup
depends on how the pump pulse interacts with and thereby resolves the
transversal structure of the probe pulse. For a transversally smooth Gaussian
pump  we discovered a smooth second-order transition in a signal-photon
observable as a function of the pump's beam waist (or equivalently the beam
waist ratio) serving as a control parameter. For a transversally structured
flat-top pump, the phase transition is of first order.

In the present example, the language of critical phenomena was also useful
to quantify the second-order transition in terms of a critical scaling law and a
corresponding critical exponent for the order parameter. In contrast to
such critical phenomena in statistical physics, we do not observe an
independence of the critical quantities from the details of the pulse collision,
and thus, there is no notion of universality classes. Still, we observe an effective
universality, \ie an insensitivity of the critical exponent to specification
details in a larger parameter region.

Another difference to conventional phase transitions is
reflected in the role and realization of the symmetry: Even though the O(2)
axisymmetry is important for the qualitative discussion of the phase transition
in the first place, the phase transition does not correspond to an
order-disorder transition nor is the symmetry spontaneously broken on one side
of the transition. This is because there is no notion of a ground state that
may or may not respect the symmetry. Instead, the observables are quantum
averages over all possible scattering states, implying that the observables on
both sides of the transition fully respect the O(2) symmetry.

As a further important remark, it should be emphasized that the choice of laser
parameters made in this work is particularly suitable for the discussion of the
phase transition analog. By contrast, the parameters are not optimized for a
realistic first-discovery experiment of quantum nonlinearities. For the latter,
an optimization of the relative polarization angles, the choice of blocking
fraction of the annular beam, and the waist size ratio of the beams would
suggest partly very different parameter regimes; see
\cite{Schutze:2024kzu,Ahmadiniaz:2024xob} for the discussion of concrete
detection schemes.

Finally, we stress again that the present work and the quantitative analysis of
the partly subtle phenomena have required the use of a reliable numerical
simulation code such as the VacEm code together with some performance
improvements as well as a careful error analysis as detailed in the appendixes.
We believe that the present work serves as a useful example for the use of such
codes as well as a motivation for their further development in the future.

\acknowledgments \label{sec:acknowledgments}

We thank Fabian Schütze for valuable discussions. The computational resources
were provided by the HPC cluster Draco of FSU Jena. This work has been funded by
the Deutsche Forschungsgemeinschaft (DFG) under Grants No. 392856280, No. 416607684,
and No. 416611371 within the Research Unit FOR2783/2.

\appendix

\section{UNDERSTANDING AND ADVANCING THE VACEM SIMULATION CODE}
\label{sec:advancements}

The control of the numerical scheme and the subsequent error analysis require a
deeper understanding of the VacEm code following
\cite{Blinne:2018nbd,maiwald2023}. There exist two modes of operation:
\texttt{explicit} and \texttt{solver}. As the names suggest, the
\texttt{explicit} mode computes the signal amplitude based on explicitly defined
$\vect{E},\, \vect{B}$ fields. The \texttt{solver} mode takes the complex
spatial electric-field profile at the focus $\vect{E}(t_0, \vect{x})$ as input
and propagates it by solving Maxwell's equations in order to provide
$\vect{E},\, \vect{B}$ at arbitrary times $t$. The Maxwell solver was developed
in \cite{blinne2019}. Note that the theory discussed in
\cref{sec:simulating_vep} assumes $\vect{E}$, $\vect{B}$ as real fields. Within
the VacEm code itself and for the present considerations regarding the Maxwell
solver, it is nevertheless convenient to treat the fields as complex; \ie for
the remainder of this section, we assume $\vect{E}$, $\vect{B}$ to be complex
and use the replacement $\vect{E} \to \Re(\vect{E})$, $\vect{B} \to
\Re(\vect{B})$ in order to make contact with \cref{sec:simulating_vep}.

The propagation is described in terms of the complex spectral amplitudes
$a_{0p}(\vect{k})$ defined by
\begin{equation} \label{eq:electromagnetic_potential}
    \vect{A}(t, \vect{x}) = \int \frac{\diff{3}{\mathrm{k}}}{(2 \pi)^3} 
    \e^{\i \vect{k} \cdot \vect{x}} \vect{\hat{A}}(t, \vect{k}) \,,
\end{equation}
\begin{equation}
    \vect{\hat{A}}(t, \vect{k}) = \e^{- \i c \mathrm{k} t} 
    \sum_{p=1}^{2} \vect{e}_p(\vect{k}) a_{0p}(\vect{k}) \,,
\end{equation}
where $\vect{A}(t, \vect{x})$ is the electromagnetic potential and
$\vect{\hat{A}}(t, \vect{k})$ its spatial Fourier transform. Any pair of
spectral amplitudes corresponds to a solution of Maxwell's equations in vacuum.
The code utilizes the radiation gauge, \ie $A^\mu = (0,\vect{A})$ and $\nabla
\cdot \vect{A} = 0$. Starting with $\vect{E}(t_0, \vect{x})$, the complex
spectral amplitudes can be constructed as 
\begin{equation} \label{eq:spectral_amplitudes}
    a_{0p}(\vect{k}) = \e^{\i c\mathrm{k} t_0} \frac{1}{\i c \mathrm{k}} 
    \vect{e}_p(\vect{k}) \cdot \vect{\hat{E}}(t_0, \vect{k}) \,.
\end{equation}
This representation is not unique; for alternatives, see \cite{Blinne:2018nbd}.
The propagated spectral amplitudes are given by
\begin{equation} \label{eq:a1_and_a2}
    a_p(t, \vect{k}) = \e^{-i c \mathrm{k} t}  a_{0p}(\vect{k}) \,,
\end{equation}
allowing us to compute the propagated $\vect{E},\, \vect{B}$ fields in
$\vect{k}$ space, 
\begin{subequations} \label{eq:Fourier_E_and_B}
    \begin{alignat}{2}
        \vect{\hat{E}}(t, \vect{k}) 
        & = \i &c \mathrm{k} \bigl[\vect{e}_1(\vect{k})
        a_1(t, \vect{k}) + \vect{e}_2(\vect{k}) a_2(t, \vect{k})\bigr] \,, 
        \label{eq:Fourier_E} \\
        \vect{\hat{B}}(t, \vect{k}) & = &\i \mathrm{k} \bigl[\vect{e}_1(\vect{k})
        a_2(t, \vect{k}) - \vect{e}_2(\vect{k}) a_1(t, \vect{k})\bigr] \,.
    \end{alignat}
\end{subequations}
An inverse Fourier transform ($\iFFT_3$) of \cref{eq:Fourier_E_and_B} then gives
us $\vect{E},\, \vect{B}$ at arbitrary times $t$. Of course, in the
\texttt{explicit} mode,
\crefrange{eq:electromagnetic_potential}{eq:Fourier_E_and_B} are not required.

These preliminary considerations regarding the Maxwell solver in combination
with \cref{sec:simulating_vep} allow us to understand the algorithmic structure
of the VacEm code in \cref{lst:pseudocode}.

\begin{lstlisting}[label={lst:pseudocode}, caption=
    {Pseudocode algorithm of the VacEm code in \texttt{solver} mode.}]
    INPUT config.ini, $\vect{E}(t_0, \vect{x})$
    $a_{01}, \, a_{02} \longleftarrow 
    \cref{eq:spectral_amplitudes}(\FFT_3(\vect{E}(t_0, \vect{x})))$
    $\mathcal{S}_\text{a}, \, \mathcal{S}_\text{b} \longleftarrow 0$
    FOR $t$ = t_start TO t_stop STEP $\Delta t$
        $a_1, \, a_2 \longleftarrow \cref{eq:a1_and_a2}(a_{01}, \, a_{02})$
        $\hat{E}_i,\, \hat{B}_i \longleftarrow 
        \cref{eq:Fourier_E_and_B}(a_1,\, a_2)$
        $E_i,\, B_i \longleftarrow \operatorname{iFFT}_3(\hat{E}_i,\,\hat{B}_i)$
        $\mathcal{F},\, \mathcal{G} \longleftarrow
        \cref{eq:relativistic_invariants} (E_i, \, B_i)$
        $Q_i \longleftarrow \cref{eq:Q}(E_i,\, B_i,\,\mathcal{F},\,\mathcal{G})$
        $\hat{Q}_i \longleftarrow \operatorname{FFT}_3(Q_i)$
        $\mathcal{S}_\text{a} \longleftarrow \mathcal{S}_\text{a} + 
        \e^{\i c k t} \vect{e}_1 \cdot \vect{\hat{Q}}$
        $\mathcal{S}_\text{b} \longleftarrow
        \mathcal{S}_\text{b} + \e^{\i c k t} \vect{e}_2 \cdot \vect{\hat{Q}}$ 
        $R_i \longleftarrow \cref{eq:R}(E_i,\, B_i,\,\mathcal{F},\,\mathcal{G})$
        $\hat{R}_i \longleftarrow \operatorname{FFT}_3(R_i)$
        $\mathcal{S}_\text{a} \longleftarrow
        \mathcal{S}_\text{a} - \e^{\i c k t} \vect{e}_2 \cdot \vect{\hat{R}}$
        $\mathcal{S}_\text{b} \longleftarrow \mathcal{S}_\text{b} +
        \e^{\i c k t} \vect{e}_1 \cdot \vect{\hat{R}}$
    END FOR 
    $\mathcal{S}_\text{a} \longleftarrow \mathcal{S}_\text{a} \Delta t$
    $\mathcal{S}_\text{b} \longleftarrow \mathcal{S}_\text{b} \Delta t$
    OUTPUT $\mathcal{S}_\text{a},\, \mathcal{S}_\text{b}$
\end{lstlisting}

The \texttt{explicit} mode features the same algorithm minus the computation of
$\vect{E},\, \vect{B}$ and without the need to provide $\vect{E}(t_0, \vect{x})$
as input. Reference \cite{Blinne:2018nbd} implemented \cref{lst:pseudocode} in
the programming language \texttt{Python 3} \cite{Python2024}.

For the present work, two improvements of the VacEm code were implemented; see
\cite{maiwald2023} for details. These are intended to reduce the computational
cost. The original VacEm code is already feature-complete within the vacuum
emission picture in the sense that it simulates the signal amplitude
\cref{eq:signal_amplitude} for arbitrary EM fields.

The first improvement allows for the use of single-precision floating-point
operations (as an alternative to \texttt{Python}'s default double precision).
Naturally, restricting the float precision to single, nearly halves the required
memory and reduces the computation time. What makes this a feature for the VacEm
code is the fact that the reduction in float precision does not have any adverse
effect on the significance of the results. The relative error for double
precision is on the order of $10^{-16}$ (52 bits for the significand, rel. error
$2^{-52}$) and for single precision $10^{-7}$ (23 bits for the significand, rel.
error $2^{-23}$). Due to accumulation effects, the actual relative error caused
by the finite float precision is typically 1 to 2 orders of magnitude
larger [$\FFT_3$ error growth is $\mathcal{O}(\log(N_x N_y N_z))$]. For our annular flat-top probe and Gaussian pump setup using single
precision, we find no (additional) error for $\varphi_\text{peak}$ and a relative
error of $10^{-5}$ for the number of signal photons in the background-free
region $N_\text{hole}$ compared to the double precision result. This precision
restriction is well below the analytical error of the one-loop approximation
around $10^{-2}$ and the total numerical error. In fact, when working on
predictions for experiments, half-precision floating-point operations could be
considered (currently not implemented and therefore not tested). The
implementation of single-precision floating-point operations for the VacEm code
is only done for the three-dimensional arrays, as all other parts are
computationally insignificant in comparison.

The second improvement is multinode parallelism. The original VacEm code
employs \texttt{pyFFTW} \cite{pyFFTW2012} to compute the $\FFT_3$ in parallel.
This implementation is restricted to one node. To reduce computation time, we
added parallelism to the time integration through \texttt{mpi4py}
\cite{mpi4py2023}. The allowed values for the number of nodes is restricted to
powers of 2. This is due to the parallel data transfer stage. The speedup is
close to ideal for a small number of nodes. There are no memory savings. In
fact, the full amount of memory required in the original VacEm implementation is
now allocated on each node; \ie the total memory usage scales by the number of
nodes plus an additional memory overhead.

As discussed in \cref{sec:numerics}, the memory per node is the main limiting
factor for the VacEm code. A lower bound for the memory usage at double
precision is given by \cite{maiwald2023}
\begin{equation} \label{eq:RAM_min}
    \begin{aligned}
        \mathrm{RAM}_\text{min} = N_x N_y N_z 
        & (15 \times \SI{128}{bit} \\
        & + 8 \times \SI{64}{bit}) \frac{10^{-9}\,{\rm GB}}{\SI{8}{bit}}\,,
    \end{aligned}
\end{equation}
since 15 \texttt{complex128} ($a_{01}$, $a_{02}$, $\mathcal{S}_\text{a}$,
$\mathcal{S}_\text{b}$, $a_1$, $a_2$, $\vect{E}$, $\vect{B}$, $\vect{Q}$ or
$\vect{R}$) and 8 \texttt{double} ($\mathcal{F}$, $\mathcal{G}$, $\vect{e}_1$,
$\vect{e}_2$) three-dimensional arrays are allocated at the same time. The only
scalable way around this problem is $\FFT_3$ with distributed-memory
parallelism. For the distributed memory allocation and $\FFT_3$ computation,
there already exist multiple libraries. Based on \texttt{MPI} and \texttt{FFTW},
there are \texttt{mpi4py-fft} \cite{mpi4py-fft2017,dalcin2018},
\texttt{fftw3-mpi} \cite{fftw2024,Frigo:2005zln}, \texttt{pfft}
\cite{PFFT2014,pippig2013}, and \texttt{p3dfft++}
\cite{P3DFFT++2017,pekurovsky2012}, all available through the \texttt{Python}
wrapper \text{FluidFFT} \cite{FluidFFT2017,mohanan2019}. For different $\FFT_3$
implementations including GPU capabilities, see \texttt{HeFFTe}
\cite{HeFFTe2021,ayala2020}. Yet, integrating a distributed memory scheme into
the current VacEm code structure is not straightforward.

In addition to these generic improvements, there is some potential to optimize
the algorithm (\cref{lst:pseudocode}) itself, \eg making use of in-place
operations.

\section{STRATEGIES FOR REDUCING NUMERICAL ARTIFACTS} \label{sec:artifacts}

The VacEm code's main operations are the spatial $\FFT_3$ and the temporal
rectangle-rule integration used to perform the required space-time integrations.
As already previously stated in \cref{sec:numerics}, the discrete Fourier
transform exposes us to aliasing and spectral leakage.

Avoiding aliasing is straightforward by sampling above the Nyquist rate,
provided that the maximum frequency in each spatial dimension is known. For the
counterpropagating axisymmetric pulse collision setups in this work, we have one
propagation direction $y$ and two transversal directions $x$, $z$. The maximum
wave number in the direction of propagation is $2 \pi / \lambda$ plus a
bandwidth term $\propto 1/(c \tau)$. In the transversal directions, we only have
to resolve wave numbers of the order of $\propto 1/w_{0}$. Thus, we obtain an
estimate of the maximum wave vector $\tilde{k}_{x,y,z}$. For the grid
resolution, we find $L_y/N_y \ll L_{x,z}/N_{x,z}$. Our spatial grid sizes at 7.5
points per wavelength range from $63 \times 637 \times 63$ to $360 \times 637
\times 360$ and are therefore comparatively lightweight. The numerical cost of
simulations without the symmetry provided by the counterpropagating axisymmetric
setup would be much larger since more than one axis needs to resolve $2 \pi /
\lambda$.

\begin{figure}
    \includegraphics[width=\columnwidth]{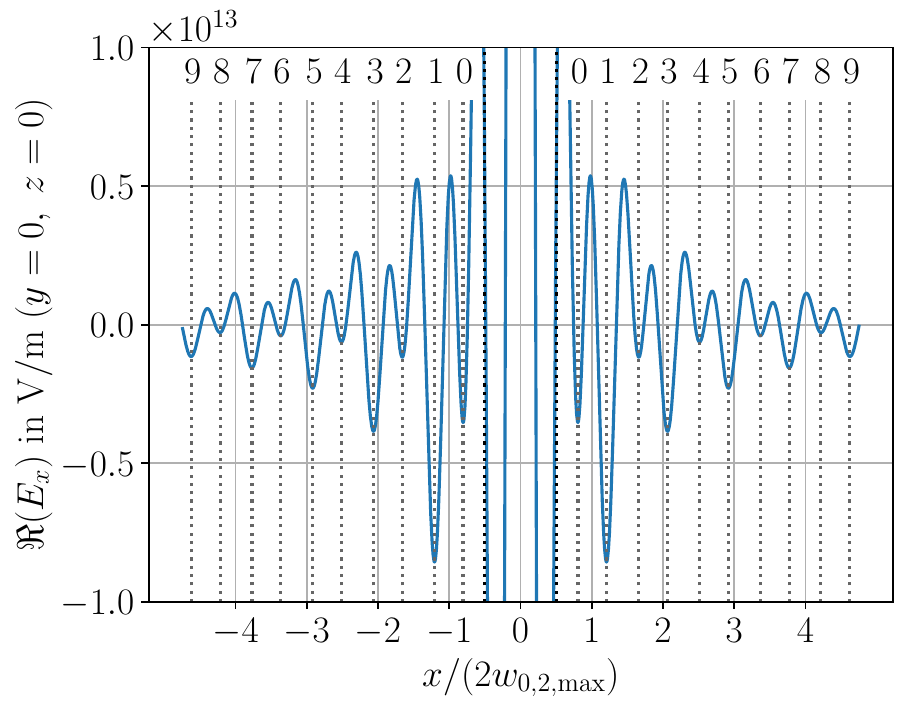}
    \caption{Spectral leakage is reduced by a suitable adjustment of the
    boundaries of the simulation domain. In the transversal directions, we
    choose the axis lengths $L_{x,z}$ such that the boundaries lie in the minima
    of the probe's focus profile. The gray dotted lines labeled from
    \numrange{0}{9} depict a sequence of corresponding spatial cutoffs used for
    our error estimate.}
    \label{fig:side_peak_cutoff}
\end{figure}

To reduce the effects of spectral leakage, the periodic continuation of the
spatial simulation domain should be as smooth as possible. We focus on the
annular flat-top probe, as it contributes linearly to the signal amplitude. The
cutoff at the boundaries of the simulation domain should be at extremal points
of the annular flat-top beam's field amplitude at $t = 0$. Along the propagation
direction, we cut off at the next local minimum after $L_y = 6 c \tau$. For the
transversal directions, we set $L_{x,z} = 2 w_{0,2,\text{max}}$ with
$w_{0,2,\text{max}} = \SI{8.0}{\micro m}$ as the lower bound and cut off at the
next 10 local minima as visualized in \cref{fig:side_peak_cutoff}. This provides
a range of $L_{x,z}$ for the error estimation in \cref{sec:error_estimation}.
In \cref{fig:side_peak_cutoff}, the black dotted lines show the lower bound, and the
gray dotted lines labeled from \numrange{0}{9} show the transverse domain
boundaries. As already mentioned, the domain is always determined for
$w_{0,2,\text{max}}$ irrespective of the actual pump beam waist. This ensures
that the probe is always identically resolved for all pump beam waists, and its
resolution is only subject to the changing extents $L_{x,z}$ of the simulation
in the transverse directions. On the other hand, for fixed $L_{x,z}$, the pump
is best resolved if its beam waist is small. At cutoff 9 used for the results
presented in \cref{sec:numerical_simulation}, this effect is of no importance
anymore.

At last, we have to consider the number of time steps needed for sufficient
convergence of the rectangle-rule integration in time. Of course, resolving the
frequency $c/\lambda$ is required. Significantly more time steps are needed for
the rectangle rule to yield a reasonable approximation. We use $N_t = 384$
corresponding to 7.5 points per period, matching the spatial resolution; see
\cite{Blinne:2018nbd} for a convergence analysis of the VacEm code's time
integration in the all-optical regime.

Note that in addition to the VacEm code's intrinsic error sources the evaluation
of the simulation data is nontrivial. Most observables relevant to experiment
are formulated in spherical coordinates. A mapping from Cartesian to spherical
coordinates via interpolation is necessary and can cause artifacts as well as
significantly contribute to the total numerical error.

\section{ERROR ESTIMATION FOR A MULTIDIMENSIONAL CONVERGENCE PROBLEM}
\label{sec:error_estimation}

\begin{figure}
    \includegraphics[width=\columnwidth]{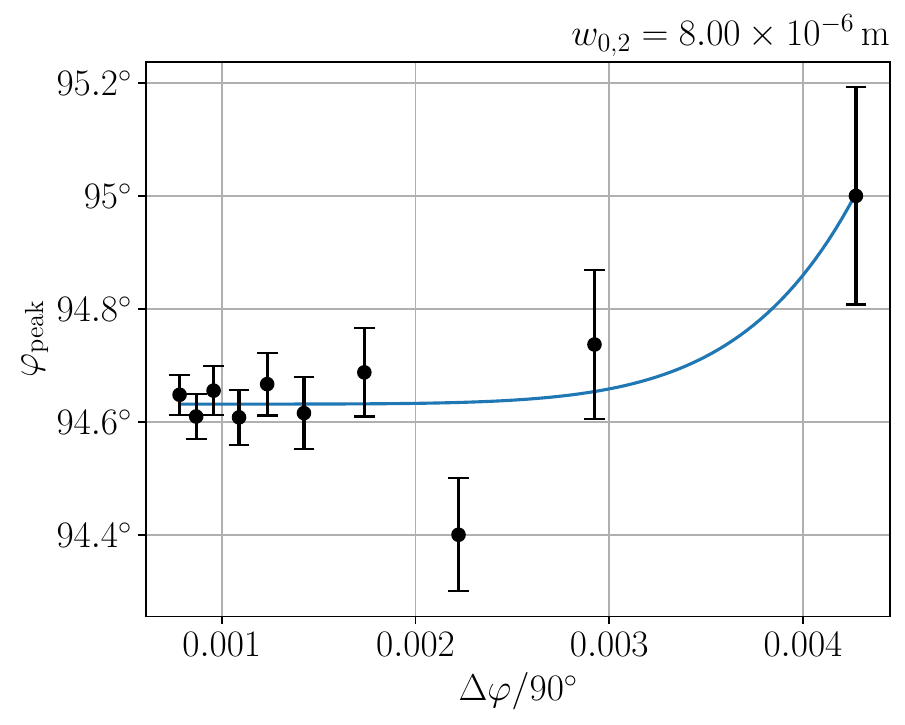}
    \caption{Convergence of $\varphi_\text{peak}$ for the annular flat-top probe
    and Gaussian pump setup at $w_{0,2} = \SI{8}{\micro\meter}$ as a function of
    the azimuthal resolution $\Delta \varphi$. The blue curve shows the fit of
    the Richardson extrapolation model \cref{eq:Richardson_extrapolation}.}
    \label{fig:vbs_xz_meta_phi_peak_9}
\end{figure}

\begin{figure}
    \includegraphics[width=\columnwidth]{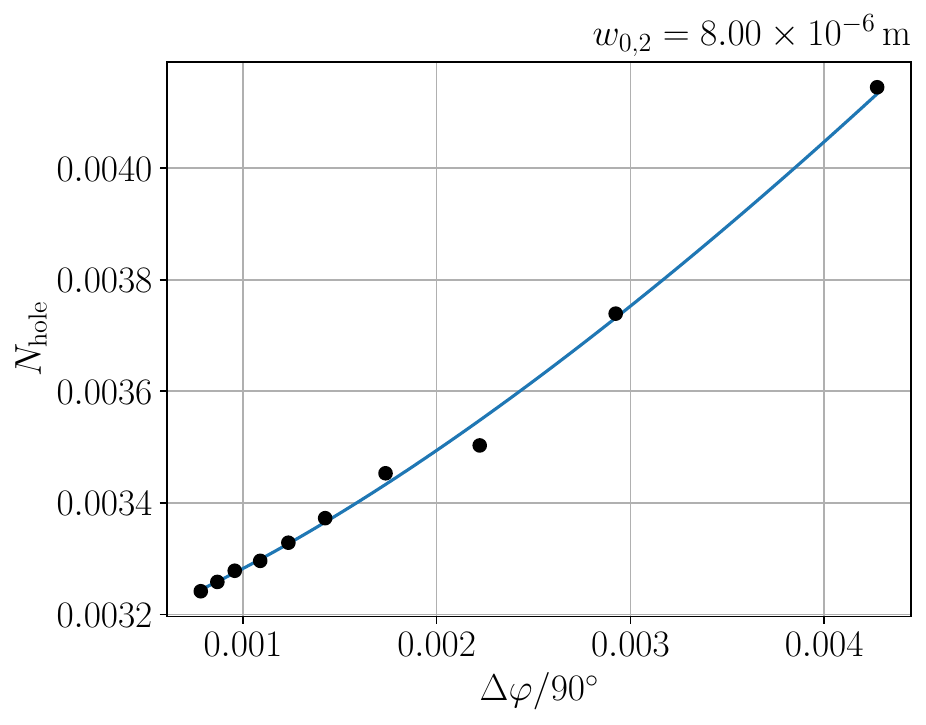}
    \caption{Convergence of $N_\text{hole}$ for the annular flat-top probe and
    Gaussian pump setup at $w_{0,2} = \SI{8}{\micro\meter}$ as a function of the
    azimuthal resolution $\Delta \varphi$. The blue curve shows the fit of the
    Richardson extrapolation model.}
    \label{fig:vbs_xz_meta_N_hole_9}
\end{figure}

There are eight parameters controlling the simulation grid of the VacEm code. For
each of the four axes spanning the simulation grid, there is one length $L_\mu$
and one number of points $N_\mu$. Convergence is expected for $L_\mu \to \infty$
and $N_\mu \to \infty$. Focusing on just one parameter, Richardson extrapolation
\cite{richardson1911} is a universal approach to determine the convergence rate
and the extrapolated true value of an observable. Due to the axial symmetry of
our setups with regard to the propagation axis ($y$ axis), we can treat the
transverse axis lengths $L_{x,z} = 2 \pi / k_{x,z} \propto 1/\Delta \varphi$ as
one parameter. Strictly speaking, the axial symmetry is broken by the
polarization which we choose to lie along the $z$ axis. Yet, an axial symmetry
is present for the individual field components, \eg $E_x$, $E_y$, $E_z$. This is
what matters for us since the $\FFT_3$ acts on these field components. For the
observable $\varphi_\text{peak}$ (now considered as a function of the
discretization) in
\cref{fig:vbs_xz_4_5_meta_phi_peak_result,fig:vbs_flat_xz_1_2_meta_phi_peak_result},
the Richardson extrapolation is given by
\begin{equation} \label{eq:Richardson_extrapolation}
    \begin{aligned}
        \varphi_\text{peak}^\ast
        & = \varphi_\text{peak}\left(\frac{\Delta \varphi}{\pi/2}\right) \\
        & \quad + C \left(\frac{\Delta \varphi}{\pi/2}\right)^p
        + \mathcal{O}\left(\left(\frac{\Delta \varphi}{\pi/2}\right)^{p+1}
        \right) \,,
    \end{aligned}
\end{equation}
where $^\ast$ marks the extrapolated true value and $p$ is the convergence rate
with regard to the discretization (azimuthal resolution or step size) $\Delta
\varphi$. At the reference scale $\pi/2$, we get $\frac{\Delta \varphi}{\pi/2}
\ll 1$, justifying a truncation at leading order. An example for the convergence
fit of the observable $\varphi_\text{peak}$ is given in
\cref{fig:vbs_xz_meta_phi_peak_9} for the case of the annular flat-top probe and
Gaussian pump setup at $w_{0,2} = \SI{8}{\micro\meter}$.
Studying the convergence of the VacEm code for different observables, an
oscillatory behavior becomes evident. The Richardson model
\cref{eq:Richardson_extrapolation} is unable to fit these oscillations but still
remains a valuable tool for extracting the overall trend, provided that
sufficiently many data points are available. In
\cref{fig:vbs_xz_meta_phi_peak_9}, oscillations dominate at low azimuthal
resolution, and therefore the convergence rate cannot be accurately determined;
\ie the relative error in $p$ is greater than 1. Adding more data points is
costly and partly constrained by our cutoff requirement discussed in
\cref{fig:side_peak_cutoff}. Nevertheless, we obtain a reasonable estimate of
the extrapolated true value; here $\varphi_\text{peak}^\ast
\big|_{\cref{fig:vbs_xz_meta_phi_peak_9}} = \SI{94.631 \pm 0.038}{\degree}$. In
view of these difficulties in model fitting, we use $\varphi_\text{peak}^\ast$
only for error estimation,
\begin{equation} \label{eq:error_estimation}
    \Delta \varphi_\text{peak} 
    = |\varphi_\text{peak} - \varphi_\text{peak}^\ast| \,.
\end{equation}
\Cref{eq:Richardson_extrapolation,eq:error_estimation} can be analogously used
for other observables and their corresponding discretization. For a less complex
and more experimentally relevant observable---the number of signal photons in the
background-free region $N_\text{hole}$---we encounter weaker oscillations and
therefore better compatibility with the Richardson model; see
\cref{fig:vbs_xz_meta_N_hole_9}.
The convergence rate of $N_\text{hole}$ can be determined as $p
\big|_{\cref{fig:vbs_xz_meta_N_hole_9}} = \num{1.38 \pm 0.21}$. The oscillatory
convergence behavior can be understood as an artifact of the discretization---primarily in the context of the $\FFT_3$ but also regarding the construction of
a given observable. Increasing $L_{x,z}$ does not just add points to the
$\vect{k}$ space grid but, in general, it readjusts the position of all points, as
we keep the grid resolution in coordinate space fixed. Thus, especially
angle-dependent observables like $\varphi_\text{peak}$ undergo significant
oscillations during convergence. Integration along the discretized axes
smoothens out this problem; \cf \cref{fig:vbs_xz_meta_N_hole_9}.

A scaling of the remaining simulation parameters gives corrections on the same
order of magnitude or smaller for all test cases that we performed. With the
annular flat-top probe and Gaussian pump setup, an increase of the grid point
density from 7.5 to 9 points per period and wavelength shows relative
corrections on the order of $10^{-2}$ for $N_\text{hole}$ and $10^{-4}$ for
$\varphi_\text{peak}$. Extending the propagation length $L_{t,y}$ by one pulse
duration $\tau$ results in $10^{-4}$ for $N_\text{hole}$ and no corrections for
$\varphi_\text{peak}$ (since the mapping to spherical coordinates is identical).
In general, each simulation parameter has its own convergence rate for a given
observable. A multidimensional convergence analysis according to the Richardson
model is given by
\begin{equation} \label{eq:multidimensional_Richardson}
    A^\ast = 
    A(d_1,\ldots,d_n) + \sum_{i = 1}^{n} \left(C_i d_i^{p_i} 
    + \mathcal{O}\left(d_i^{p_i + 1}\right)\right) \,,
\end{equation}
with observable $A$, number of simulation parameters $n$, discretizations $d_i$,
convergence rates $p_i$, extrapolated true value $A^\ast$, and leading-order
constants $C_i$. Unfortunately, due to the large number ($2n+1$) of model
parameters $A^\ast$, $p_i$, $C_i$, fitting \cref{eq:multidimensional_Richardson}
is currently not feasible.

\begin{figure}
    \includegraphics[width=\columnwidth]{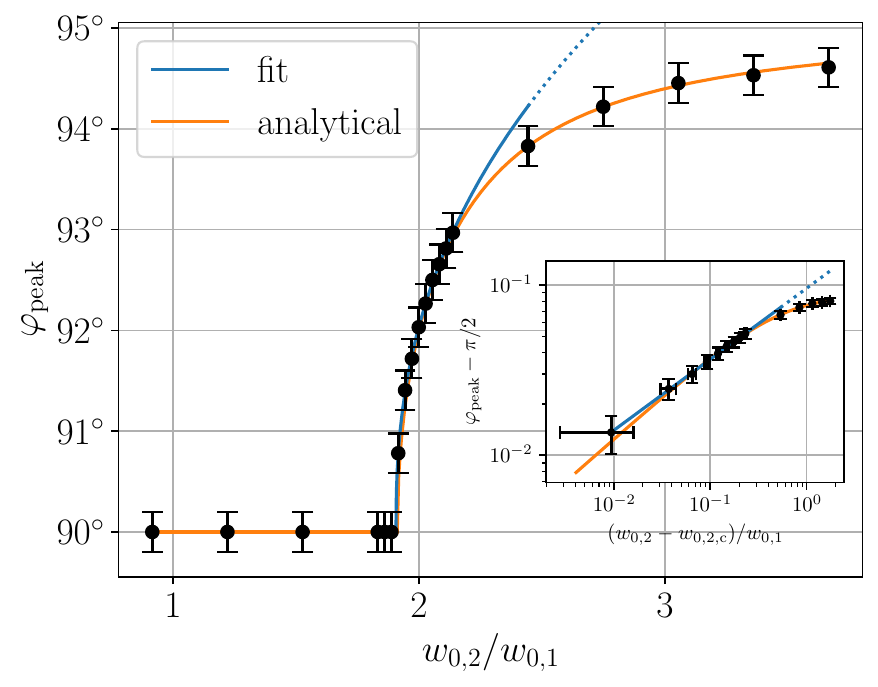}
    \caption{Main emission direction of the far-field signal-photon distribution
    for an annular flat-top probe colliding with a Gaussian pump as a function
    of the pump-to-probe waist ratio $w_{0,2}/w_{0,1}$ across the transition.
    The figure is completely analogous to
    \cref{fig:vbs_xz_4_5_meta_phi_peak_result} but accounts for errors from all
    8 discretization parameters using \cref{eq:total_discretization_error} and a
    linear convergence rate $p=1$. This yields more conservative estimates for
    the critical point $(w_{0,2,\text{c}}/w_{0,1})\big|_\text{g} = \num{1.9065
    \pm 0.0066}$ and the critical exponent $\boldsymbol{\beta} = \num{0.417 \pm
    0.083 }$, in satisfactory agreement with both the simulation result and the
    analytical estimate given in the main text.}
    \label{fig:vbs_xz_4_5_total_error_meta_phi_peak_result}
\end{figure}

Still, \cref{eq:multidimensional_Richardson} can be used to obtain an estimate
for the total discretization error in all 8 simulation grid parameters of the
VacEm code. The error for a given observable $\Delta A(d_1/s, \ldots, d_n/s)$ at
discretizations $d_i/s$, where $s>1$ is an arbitrary scaling factor with respect
to some base discretization $d_i$, can be simplified to 
\begin{equation} \label{eq:total_discretization_error}
    \begin{aligned}
        \Delta A(d_1/s, &\ldots, d_n/s) \\ 
        = \Bigg| &\frac{A(d_1/s, \ldots, d_n/s) - A(d_1, \ldots, d_n)}{s^p - 1} 
        \Bigg| \,,
    \end{aligned}
\end{equation}
under the assumption of an effective convergence rate $p_i = p$. Using $L_t =
\tau$, $L_y = 2 c \tau$, $L_{x,z} = 10 w_{0,2,\text{max}}$, a temporal
resolution of 6 points per period $\lambda/c$, and a spatial resolution of 3
points per wavelength $\tilde{k}_{x,y,z}$ as the base discretization before
scaling with $s = 2$, we simulate the collision of the annular flat-top probe and
Gaussian pump again, now estimating the total discretization error
\cref{eq:total_discretization_error} for linear convergence $p = 1$. The
resulting continuous phase transition in
\cref{fig:vbs_xz_4_5_total_error_meta_phi_peak_result} matches our findings
(\cref{fig:vbs_xz_4_5_meta_phi_peak_result}) from the main text.
This estimate is reliable and acts as an upper bound provided that the VacEm
code's effective convergence rate $p_\text{eff} \geq 1$. This assumption is
plausible but needs to be verified case by case.

For $N_\text{hole}$, we find a relative error between \SI{7}{\percent} and \SI{15}{\percent}.
In the case of $\varphi_\text{peak}$, the discretization error in $\varphi$ due
to the mapping from the Cartesian grids at discretizations $d_i$ and $d_i/2$ to
spherical coordinates exceeds the estimated total discretization error (which
does not take the spherical mapping into account). Therefore, the errors shown
in \cref{fig:vbs_xz_4_5_total_error_meta_phi_peak_result} are the sum of the
$\Delta \varphi$ at both discretizations.

\FloatBarrier
\bibliography{paper1.bib}

\end{document}